\documentclass{article}

\usepackage{arxiv}

\usepackage[utf8]{inputenc} 
\usepackage[T1]{fontenc}    
\usepackage{hyperref}       
\usepackage{url}            
\usepackage{booktabs}       
\usepackage{amsfonts}       
\usepackage{nicefrac}       
\usepackage{microtype}      
\usepackage{graphicx}
\usepackage{tabularx}
\usepackage{float}
\usepackage{amsmath}
\usepackage{threeparttable}
\usepackage{caption}

\title{When the Environment Becomes the Interface: Multisensory Environmental Interfaces for Human-AI Interaction in Autonomous Vehicles}
\date{September 2026}

\hypersetup{
  pdftitle={When the Environment Becomes the Interface: Multisensory Environmental Interfaces for Human-AI Interaction in Autonomous Vehicles},
  pdfauthor={Keqi Chen, Runjia Tan, Xinyi Fu, Shanhe Lou, Kwan Min Lee, and Chen Lv},
  pdfkeywords={autonomous vehicles, environmental interface, atmospheric design, multisensory interaction, passenger experience, human-AI interaction, technology trust, S-O-R framework},
  hidelinks
}

\author{
 Keqi Chen \\
  School of Mechanical and Aerospace Engineering\\
  Nanyang Technological University\\
  Singapore, SG 639798 \\
  \texttt{keqi001@e.ntu.edu.sg} \\
  \And
 Runjia Tan \\
  School of Mechanical and Aerospace Engineering\\
  Nanyang Technological University\\
  Singapore, SG 639798 \\
  \texttt{runjia.tan@ntu.edu.sg} \\
  \And
Xinyi Fu \\
  The Future Laboratory\\
  Tsinghua University\\
  Beijing, China\\
  \texttt{fuxy@mail.tsinghua.edu.cn} \\
  \And
Shanhe Lou \\
  School of Mechanical and Aerospace Engineering\\
  Nanyang Technological University\\
  Singapore, SG 639798 \\
  \texttt{shanhe.lou@ntu.edu.sg} \\
  \And
 Lee Kwan Min\\
  Wee Kim Wee School of Communication and Information\\
  Nanyang Technological University\\
  Singapore, SG 639798\\
  \texttt{kwanminlee@ntu.edu.sg}\\
  \And
 Chen Lv\\
  School of Mechanical and Aerospace Engineering\\
  Nanyang Technological University\\
  Singapore, SG 639798 \\
  \texttt{lyuchen@ntu.edu.sg} \\
}

\begin{document}
\maketitle
\begin{abstract}

As artificial intelligence increasingly assumes operational control, human-computer interaction is shifting from operating systems through explicit interfaces toward inhabiting intelligent environments. This transformation raises a fundamental question: when users no longer directly manipulate a system, what becomes the interface between humans and intelligent technologies? This study introduces the concept of environmental interfaces---designed environmental conditions that mediate human-AI relationships through ambient, holistic, and evaluative pathways rather than explicit functional interactions. Using autonomous vehicle cabins as a theoretically revealing context, we conducted a within-subject experiment with 24 participants across 216 observations, manipulating lighting and scent conditions in a simulated autonomous driving environment. Results reveal three key insights. First, environmental perception emerged as a primary interaction pathway, accounting for 66.5\% of the variance in journey experience evaluation, demonstrating that environmental conditions can function as an interface rather than merely a supportive design element. Second, multisensory processing followed a hierarchical architecture: individual sensory appraisals were initially processed independently, while cross-modal integration emerged selectively during higher-order environmental evaluation rather than at early perceptual stages. Third, olfactory stimuli exerted stronger effects than visual stimuli on affective responses and experience evaluation, challenging the conventional visual dominance in automotive interaction design. Furthermore, environmental quality and sensory congruency significantly predicted trust in autonomous systems, suggesting that users may rely on environmental cues as proxy signals when direct assessment of AI competence is unavailable. These findings extend human-AI interaction theory by identifying environmental interfaces as a distinct interaction modality and highlight a broader transition from designing interfaces for operating intelligent systems toward designing environments for human inhabitation of intelligent systems.

\end{abstract}


\section{Introduction}
The transition to autonomous vehicles fundamentally transforms human-vehicle interaction, shifting occupants from operators engaged in active control to passengers inhabiting intelligent mobile environments \cite{-_Kun2016-tw, -_Wintersberger2021-ri}. This transformation liberates cognitive resources previously dedicated to driving tasks, enabling vehicles to function as spaces for diverse activities including work, relaxation, and social interaction \cite{-_Fagnant2015-af, -_Pettigrew2023-he}. As operational control transitions to AI systems, the human-vehicle relationship evolves from tool manipulation to environmental cohabitation, fundamentally altering design priorities for in-cabin experiences.

However, current automotive HCI research remains largely anchored in paradigms developed for driver-operated vehicles, where decades of investigation have emphasized distraction avoidance and attentional segregation to protect driver focus on safety-critical tasks \cite{-_Strayer2001-gn, -_Spence2015-dp, -_Khan2020-hi, -_Detjen2021-qm}. While appropriate for operational contexts, these frameworks prioritize functional support over experiential enhancement. When passengers are freed from driving responsibilities, the cabin environment transitions from a potential source of distraction to a primary locus of experiential attention \cite{-_Wintersberger2019-cl,Dam2025-zq}. This paradigm shift necessitates theoretical frameworks for passenger-centric design that prioritize experience quality over distraction avoidance \cite{-_Obrist2017-ux,Dani2019-fj}.

More fundamentally, this transformation represents a shift from \textit{operating} systems through interfaces to \textit{inhabiting} environments with systems. In conventional vehicles, interaction occurs through explicit control interfaces (steering wheels, pedals, switches) that mediate the human-vehicle relationship. These physical interfaces structure engagement: they are the touchpoints through which drivers manipulate systems and receive continuous feedback. Autonomous vehicles eliminate this interface layer. Passengers no longer manipulate controls or monitor displays; they simply occupy a space that moves them. Yet this does not mean interaction disappears---it transforms. Passengers still need ways to experience the AI system, form trust in its capabilities, and evaluate journey quality. The question becomes: \textit{when explicit functional interfaces vanish, what mediates the passenger-system relationship?}

We propose that when operational controls are removed or diminished in salience, the designed cabin atmosphere itself functions as an interface---what we term an \textit{environmental interface}. In this conceptualization, environmental quality becomes a medium through which passengers experience and potentially evaluate autonomous systems. Unlike traditional functional interfaces supporting explicit manipulation, environmental interfaces operate through ambient, holistic mechanisms. Environmental quality may serve as a tangible proxy for system sophistication when direct performance assessment proves constrained \cite{Spence2020-bu}, with potential relevance for human-AI interaction more broadly, particularly in contexts where explicit control interfaces are absent or backgrounded \cite{Weiser1991-oy, Abowd2000-vw}.

The concept of \textit{atmospherics}---the intentional design of multisensory environments to shape holistic experience---offers theoretical grounding for environmental interfaces. Atmospheric design principles have demonstrated robust effects in commercial environments, influencing behavior, satisfaction, and environmental evaluation through coordinated sensory interventions \cite{Kotler1973-es, Baker2002-lk, Morrison2011-uq, Turley2000-ch}. The S-O-R framework underlying atmospheric research provides conceptual structure for understanding how environmental stimuli may influence internal psychological states, which in turn drive behavioral and evaluative responses \cite{Mehrabian1974-cg}. However, atmospheric research has developed primarily in commercial environments such as retail, hospitality, and healthcare settings. Autonomous vehicle cabins represent a distinct environmental context with different characteristics: passengers inhabit a mobile, technology-mediated space rather than engaging with a stationary commercial setting. Whether atmospheric principles documented in commercial environments translate effectively to autonomous vehicle contexts, and through what psychological mechanisms, remains an open empirical question requiring systematic investigation \cite{Spence2014-jn}.

Despite theoretical importance, systematic investigation of environmental factors in autonomous vehicle passenger experience remains limited \cite{Zhang2025-gh,Nikitas2021-mr}. This study addresses this gap through systematic empirical examination of multisensory atmospheric effects in autonomous mobility contexts. Our investigation is guided by two central research questions:

\textbf{RQ1}: To what extent do multisensory atmospheric stimuli influence passenger experience across multiple psychological dimensions including sensory evaluations, environmental perceptions, trust formation, and experiential outcomes?

\textbf{RQ2}: What psychological processes mediate the relationships between multisensory environmental stimuli and passenger outcomes, and do these mechanisms operate through sequential processing stages as suggested by environmental psychology theory?

By addressing these questions through systematic multisensory manipulation in an immersive autonomous vehicle simulation, this study develops an empirically-grounded model of environmental interfaces in passenger experience. Our findings reveal the psychological architecture through which atmospheric design influences not only passenger comfort but also fundamental evaluations of journey quality and technology acceptance, offering theoretical insights for human-AI interaction in contexts where explicit control interfaces are absent or diminished in salience.

\section{Related Work}

\subsection{The Driver-Centric Legacy and Its Constraints}

Automotive HCI has long operated within what we term the \textit{operational support paradigm}, a research framework prioritizing driver performance, safety, and distraction minimization \cite{Engstrom2017-yi, Wickens2013-dg}. This paradigm emerged logically from driving's safety-critical nature, treating the vehicle cabin as a cockpit where any non-essential stimulus poses potential liability. While this approach served driver-operated vehicles well, it systematically constrained how researchers conceptualize in-cabin sensory environments. Examining sensory modality research reveals this constraint's breadth. Auditory research focused on alert design and infotainment cognitive load rather than acoustic quality \cite{Spence2017-ku}. Haptic research examined operational feedback through steering wheel vibrations or seat-based warnings, treating touch as another alert channel \cite{Hogema2011-ji}. Automotive lighting research concentrated almost exclusively on alertness optimization and circadian regulation to combat fatigue \cite{Cajochen2007-xe, Bullough2021-vl}. When ambient systems were introduced, researchers primarily validated non-interference, ensuring atmospheric elements did not compromise driving tasks \cite{Reisinger2019-sb}. Olfactory research pursued similarly utilitarian goals: scent-based warnings and vigilance maintenance \cite{Ho2017-fj, Dmitrenko2017-gq}. Spence and Ho \cite{Ho2017-fj} term the paradigm's organizing principle \textit{attentional segregation}: deliberate isolation of sensory inputs to prevent interference with primary driving tasks. Decades of driver distraction research reinforced this imperative, with studies demonstrating that integrated or complex sensory experiences can impair performance \cite{Strayer2001-ff}. This logic proved sound when humans controlled vehicles but becomes questionable as AI assumes this role.

Early autonomous vehicle research largely extended driver-centric assumptions. Studies focused on transitional automation (SAE Level 3), treating passengers as fallback operators who must maintain situational awareness for potential takeovers \cite{Kyriakidis2015-hm, Gold2016-jm, Wintersberger2019-eu}. Research priorities centered on takeover interface design, situational awareness maintenance, and motion sickness mitigation \cite{Dam2025-zq}. While these concerns remain valid for transitional automation, they overlook experiential considerations relevant to full automation (SAE Levels 4-5), where passengers never need to resume control. 

\subsubsection{Passenger Experience Research: Current Landscape and Limitations}

Recent autonomous vehicle research has begun addressing passenger experience, though through approaches that leave atmospheric mechanisms largely unexplored. Studies cluster in two categories: acceptance-focused investigations and multisensory functional applications. 

Acceptance-focused research examines psychological and functional factors influencing AV adoption. Zhang et al. \cite{Zhang2025-gh} systematically reviewed passenger immersive experiences, identifying emotional and sensory factors alongside interaction, trust, and dispositional dimensions. Kolarova and Cherchi \cite{Kolarova2021-hs} demonstrated that travel experience quality significantly influences autonomous vehicle value perceptions. Nastjuk et al. \cite{Nastjuk2020-jb} identified experiential factors beyond functional acceptance as critical for adoption. While these studies establish experience as important, they treat it as an outcome variable explained by technology attributes (perceived usefulness, trust, social influence) rather than investigating the environmental mechanisms that shape it. 

Multisensory functional applications examine sensory stimuli for specific performance objectives. Research addresses system communication and alerting \cite{dmitrenko2018smell, locken2020increasing}, motion sickness mitigation \cite{schartmuller2020sick}, and physiological stress reduction \cite{wang2025vehicle}. These investigations document discrete sensory effects on functional outcomes, including improved response times, alertness management, and stress recovery, without examining how integrated multisensory environments function as holistic atmospheres shaping passenger experience. Studies focus on physiological or behavioral responses without systematically investigating psychological processing pathways through which multisensory atmospheres might influence passenger psychology.

Both research streams make important contributions while leaving critical gaps. Acceptance research identifies experience as important but does not specify environmental mechanisms creating it. Multisensory research demonstrates functional effectiveness of sensory modalities but does not investigate atmospheric experience or psychological processing pathways. As Zhang et al. explicitly note, "sensory feedback mechanisms and aesthetic design remain underexplored" despite recognition of sensory factors' importance \cite{Zhang2025-gh}. The question of how cabin design influences passenger experience through atmospheric channels, and through what psychological mechanisms these effects operate, remains empirically unaddressed. Understanding these mechanisms requires moving beyond operational and functional HCI toward frameworks equipped for holistic environmental experience design.

\subsection{Atmospheric Design: A Framework for Environmental Experience}

Environmental psychology offers theoretical frameworks for understanding holistic human experiences through \textit{atmospherics}, defined as intentional orchestration of environmental stimuli to influence cognitive and affective responses. Kotler \cite{Kotler1973-es} articulated the central proposition: physical environments function as active media shaping internal states and, consequently, behavior. The S-O-R framework \cite{Mehrabian1974-cg} provides conceptual structure for these processes. Environmental cues (Stimulus) influence individuals' internal emotional and cognitive states (Organism), which drive behavioral responses (Response). The framework has guided substantial research across multiple domains, though its operation in novel contexts requires empirical investigation. Studies across retail (Donovan \& Rossiter, 1982; Spangenberg et al., 2005), healthcare (Ulrich et al., 2008; Dijkstra et al., 2006), workplace (Mehta et al., 2012), and educational settings (Tanner, 2009) document systematic relationships between atmospheric interventions and human responses \cite{Donovan1982-fd, Spangenberg2005-zz, Ulrich2008-sc, Dijkstra2006-fr, Mehta2012-ss, Tanner2009-it}. In retail environments, manipulations of music, ambient scent, and lighting associate with changes in pleasure and arousal states, subsequently affecting in-store time and spending patterns. Healthcare studies show associations between environmental improvements (better lighting, nature views, reduced noise) and positive patient outcomes including accelerated recovery and reduced pain medication demands. Similar patterns emerge in workplace and educational contexts.

These cross-domain findings suggest atmospheric principles may reflect fundamental human-environment relationships rather than context-specific phenomena. However, existing evidence comes predominantly from commercial environments (retail stores, restaurants, hotels, healthcare facilities) characterized by stationary spaces and established atmospheric design practices. Autonomous vehicle cabins represent a distinct environmental context: mobile, technology-mediated spaces where passengers inhabit intelligent environments during travel. Given these contextual differences, whether atmospheric principles and their underlying psychological mechanisms documented in commercial settings operate effectively in autonomous vehicle contexts remains an open empirical question. Systematic investigation is needed to determine whether the S-O-R framework and its constituent mechanisms---sensory processing, affective responses, and evaluative outcomes---function similarly in this novel environmental context.

\subsection{Psychological Mechanisms of Environmental Experience}

Systematic investigation of atmospheric effects reveals that environmental influence operates through complex psychological mediating processes rather than direct stimulus-response relationships. Understanding these mechanisms proves essential for designing effective atmospheric interventions, particularly in novel contexts where traditional assumptions about environmental processing may not hold.

\subsubsection{Environmental Stimuli and Multisensory Integration}

Different sensory modalities function as distinct psychological triggers with varying effectiveness across contexts. Visual interventions through lighting, color, and spatial design demonstrate systematic relationships with mood, arousal, and environmental evaluation \cite{Baker2002-lk, Quartier2014-tl}. Auditory approaches using music and soundscaping show consistent effects on temporal perception and emotional states \cite{Milliman1982-rn, Roschk2017-er}. Olfactory interventions via ambient scenting prove particularly potent due to direct neurological pathways to emotional processing centers, with meta-analytic evidence showing consistent pleasure and satisfaction enhancement \cite{Spangenberg1996-zc, Roschk2017-er}. Tactile and thermal factors additionally contribute to overall atmosphere \cite{Mehta2012-ss, Peck2003-si}.

For autonomous vehicle applications, selecting appropriate sensory modalities requires consideration of the mobile, confined cabin environment and the diverse passenger activities (working, reading, resting) that may occur during travel. This context favors ambient stimuli that shape experience without demanding focal attention. Additionally, confined mobile environments require modalities that remain effective during prolonged exposure without causing adaptation or discomfort. Lighting and scent emerge as strategically optimal for autonomous vehicle applications based on two characteristics. First, both function primarily as ambient rather than focal stimuli \cite{Spence2020-bu, de2021unconscious}, capable of shaping experience from the periphery of conscious attention while exerting substantial psychological influence \cite{boyce2003human, Herz2009-rn}. This ambient quality allows them to operate without interfering with passenger activities. Second, they offer complementary psychological pathways. Lighting exerts broad cognitive and affective influence, regulating states like alertness and relaxation through circadian and attentional mechanisms \cite{kim2021emotional, blankenbach2021evaluation}. In contrast, scent provides direct access to the limbic system, centers for emotion and memory processing, enabling powerful subconscious mood regulation \cite{Herz2009-rn, dmitrenko2020caroma}. This dual-pathway approach enables robust atmospheric effects across different passenger activities and states \cite{wang2025vehicle}. Other sensory modalities present practical limitations in AV contexts. Auditory stimuli (e.g., music) conflict with passengers' diverse activity needs and individual preferences. Haptic interventions require physical contact that may feel intrusive during prolonged exposure. Thermal control, while important for comfort, operates primarily through maintenance rather than experiential enhancement. These constraints make lighting and scent particularly suitable for systematic investigation of atmospheric mechanisms.

\subsubsection{Psychological Processing Pathways}

Contemporary atmospheric theory conceptualizes the transformation of environmental stimuli into meaningful experience as a sophisticated cascade of cognitive and affective processes unfolding in hierarchical stages. Understanding this cascade proves essential for designing effective environmental interfaces.

The process begins with \textit{initial sensory appraisals}, rapid and direct hedonic judgments of individual stimuli. Zajonc \cite{Zajonc1980-ct} demonstrated that these appraisals activate reward-processing brain regions within milliseconds of exposure, occurring largely outside conscious awareness. Biederman and Vessel \cite{Biederman2006-fp} further showed that these rapid evaluations serve as foundational inputs for more complex processing. As multiple stimuli present simultaneously, the brain engages in \textit{cross-modal integration}, combining unimodal inputs to form multimodal hedonic judgments about the sensory blend \cite{Spence2011-pn}. This integration process determines whether different sensory channels reinforce or interfere with each other. Following these perceptual and integrative stages, individuals engage in \textit{higher-order environmental appraisal}, an active cognitive process extracting deeper meaning from the multisensory array \cite{gibson2014ecological}. Psychologists distinguish two core dimensions in this appraisal stage. First, \textit{sensory congruency} evaluation involves cognitive judgment about harmonious alignment of environmental elements across modalities. Mattila and Wirtz \cite{Mattila2001-js} and Spangenberg et al. \cite{Spangenberg2005-zz} demonstrated that congruent combinations facilitate fluent cognitive processing and enhance environmental evaluation, while incongruent combinations create processing conflicts diminishing atmospheric effectiveness. Second, individuals form \textit{holistic gestalt evaluations} of overall perceived atmosphere, assessing the emergent experience created by the environment rather than isolated sensory elements. Research identifies core atmospheric perception dimensions demonstrating stability across contexts: pleasantness, comfort, and aesthetic appeal \cite{Baker2002-lk, Rayburn2013-ws}. Russell and Mehrabian \cite{Russell1977-yu} suggest that these conscious environmental appraisals precede and influence final emotional experience, positioning cognitive evaluation as primary rather than secondary in the atmospheric response sequence. This proposed ordering has important design implications, suggesting that passengers may first cognitively evaluate cabin atmosphere before experiencing its emotional effects. Whether this sequence holds in autonomous vehicle contexts, where environmental quality may signal system competence in addition to providing sensory pleasure, remains an empirical question our study addresses.

\subsubsection{Outcomes in Autonomous Mobility Contexts}

This processing cascade, from sensory appraisal to configural evaluation, constitutes the Organism component of the S-O-R framework. Completing the model requires linking these internal states to relevant and context-specific outcomes (the Response component). Applying atmospheric research to autonomous mobility demands reconceptualization of outcome measures beyond traditional commercial metrics.

\textit{Holistic journey evaluation} emerges as a critical atmospheric outcome reflecting transportation's shift from instrumental activity to experiential consumption. This construct encompasses passengers' integrated assessment of comfort, enjoyment, and overall experience quality. Recent work by Kolarova and Cherchi \cite{Kolarova2021-hs} and Nastjuk et al. \cite{Nastjuk2020-jb} demonstrates that these experiential dimensions prove increasingly central to autonomous vehicle acceptance, suggesting atmospheric design may exercise disproportionate influence through its capacity to create experiential anchors organizing evaluation of all journey components.

\textit{Trust formation} represents a second critical outcome reflecting autonomous mobility's unique challenges. Lee and See \cite{Lee2004-uz} identify algorithmic opacity as creating an evaluation gap whereby passengers cannot directly assess system competence through behavioral observation. Hoff and Bashir \cite{Hoff2015-on} propose that environmental design may serve as a trust-building mechanism through competence signaling, using atmospheric quality as proxy indicator of underlying system sophistication. Baker et al. \cite{Baker2002-lk} demonstrate that high-quality environmental design systematically enhances perceptions of provider competence across contexts, with potential amplification in safety-critical transportation environments where trust assumes heightened psychological importance. Whether cabin atmosphere indeed functions as a trust signal in autonomous vehicles, and through what psychological pathways, remains empirically unexamined.

\subsection{Research Gaps and the Present Study}

The preceding review reveals that while atmospheric design principles enjoy robust empirical support across stationary commercial environments, their application to autonomous vehicle passenger experience remains largely unexplored. Three specific gaps constrain theoretical understanding and practical implementation.

First, existing autonomous vehicle research addresses passenger experience through two approaches that leave atmospheric mechanisms unexplored. Acceptance-focused studies \cite{Nastjuk2020-jb, Kolarova2021-hs, Zhang2025-gh} treat experience as an outcome variable explained by technology attributes (perceived usefulness, trust) rather than investigating environmental mechanisms that shape it. Multisensory research examines functional applications such as system alerting \cite{dmitrenko2018smell, locken2020increasing}, motion sickness mitigation \cite{schartmuller2020sick}, stress reduction \cite{wang2025vehicle}, documenting discrete sensory effects on specific outcomes without investigating how integrated multisensory environments function as holistic atmospheres. While these studies establish that experience matters and that sensory modalities can influence functional performance, the psychological mechanisms through which cabin atmospheric design influences integrated passenger experience remain unspecified. How do passengers process multisensory environmental stimuli? Through what psychological pathways do atmospheric conditions translate into experiential outcomes? These foundational questions remain empirically unaddressed despite their centrality to designing effective passenger-centric autonomous vehicle environments.

Second, atmospheric research has developed detailed models of 
psychological processing pathways (sensory appraisal, cross-modal integration, higher-order evaluation) based primarily on evidence from commercial environments such as retail, hospitality, and healthcare settings. Whether these processing mechanisms operate similarly in autonomous vehicle contexts remains an open empirical question. Autonomous vehicles present a distinct environmental context---mobile, 
technology-mediated spaces where passengers inhabit intelligent environments during travel. The extent to which atmospheric mechanisms documented in commercial settings translate to autonomous vehicle contexts, and through what specific psychological pathways atmospheric effects occur in passenger experience, requires systematic empirical investigation. Understanding whether the S-O-R processing cascade---from sensory appraisal through environmental evaluation to experiential outcomes---functions similarly across these different environmental contexts is essential for establishing theoretical foundations of atmospheric design in autonomous mobility.

Third, traditional atmospheric research emphasizes experiential outcomes such as pleasure, arousal, and satisfaction, with commercial behaviors (spending, dwell time) as ultimate dependent variables. Autonomous mobility demands different outcome conceptualizations. Beyond immediate journey satisfaction, atmospheric design may influence \textit{technology trust} when passengers cannot directly assess AI system competence through behavioral observation. Whether and how cabin atmosphere functions as an environmental pathway to trust, particularly through what psychological mediating processes, remains empirically unaddressed. This represents a critical gap given trust's central role in autonomous vehicle acceptance.

The present study addresses these gaps through systematic experimental manipulation of atmospheric stimuli (lighting and scent) in a simulated autonomous vehicle environment. We examine both the existence and the psychological architecture of atmospheric effects in passenger contexts. Specifically, we investigate: (1) whether multisensory atmospheric manipulations influence passenger experience across multiple psychological dimensions; and (2) what psychological processes mediate these relationships, particularly whether they unfold through the hierarchical processing stages suggested by environmental psychology theory. By empirically testing the S-O-R framework's internal mechanisms in autonomous mobility contexts, this study establishes foundational understanding of environmental interfaces as a distinct modality for human-AI interaction.

\section{Methodology} 
We examined multisensory atmospheric effects in autonomous vehicle passenger experience through a $3 \times 3$ within-subjects experiment manipulating lighting (blue, orange, none) and scent (mint, lavender, none). Twenty-four participants experienced all nine conditions in an immersive VR autonomous vehicle simulation, rating their experience across seven psychological dimensions. Analysis proceeded in two stages: repeated-measures ANOVA tested direct atmospheric effects (RQ1), followed by pathway modeling to examine mediating psychological processes (RQ2).

\subsection{Pilot Study and Stimulus Calibration}
Prior to the main experiment, we conducted a pilot study to determine optimal scent presentation parameters following established olfactory research protocols \cite{Herz2009-rn, Sela2010-qy}. Previous studies demonstrate that both concentration and physical distance significantly influence scent perception and hedonic response \cite{Ferdenzi2014-gk}. Ten participants (not involved in the main study) were blindfolded and exposed to essential oils at two dilution ratios (1:50 and 1:100) presented at three distances (2\,cm, 4\,cm, 6\,cm). Participants rated scent intensity and pleasantness on 7-point Likert scales after each exposure, with adequate intervals between trials to prevent olfactory fatigue \cite{Lombion-Pouthier2006-cs}. Based on pilot results, we selected 1:50 dilution ratio at 4\,cm distance, providing sufficient olfactory stimulation while avoiding overwhelming intensity.

\subsection{Participants and Recruitment}
Twenty-four participants (14 female, 10 male; age $M=29.6$, $SD=5.8$, range 23-46) were recruited through convenience sampling. All held valid driving licenses and reported normal vision and olfaction. Power analysis based on multisensory research reporting medium-to-large effects ($\eta^2_p \approx 0.15$) \cite{velasco2018multisensory, wilson2006cortical} indicated 24 participants provide .80 power to detect $f \geq 0.25$ in repeated-measures designs. The design yielded 216 observations (24 participants $\times$ 9 conditions). Observed effect sizes for significant findings ($\eta^2_p$ = .114 to .320) confirmed adequate power \cite{simner2006synaesthesia, velasco2016crossmodal}.

\subsection{Experimental Design and Apparatus}

\begin{figure}[htbp]
\centering
\includegraphics[width=.9\linewidth]{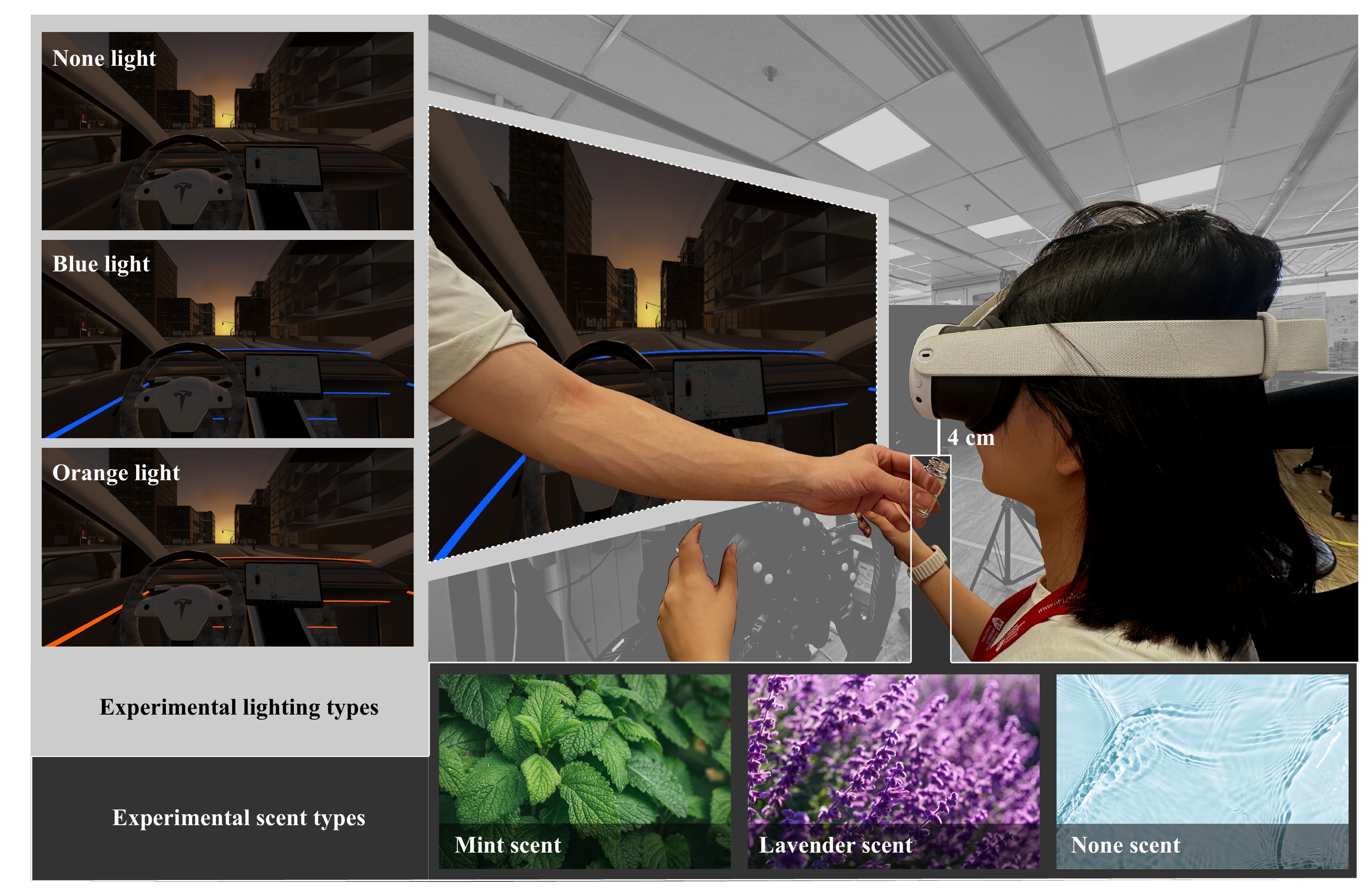}
\caption{ The virtual reality (VR) experimental environment and stimuli.}
\label{fig:experimental environment}
\end{figure}

\subsubsection{Study Design}
We employed a 3 (Lighting: blue, orange, none) $\times$ 3 (Scent: mint, lavender, none) within-subjects factorial design. This approach follows established multisensory research protocols \cite{Spence2014-jn, Velasco2021-hf}, enabling direct comparison of unimodal and multimodal effects while controlling for individual differences. We counterbalanced the presentation order of nine experimental conditions across participants using a Latin square design to mitigate order and carryover effects.

\subsubsection{Apparatus and Stimuli}
The experiment took place in a quiet, well-ventilated room ($5\,m \times 4\,m$) with controlled temperature ($22\pm1\,^\circ\mathrm{C}$) and humidity (45--55\%) to ensure consistent olfactory perception \cite{Sela2010-qy}. We simulated an autonomous urban driving experience through an immersive virtual reality (VR) environment developed in Unity. The simulation replicated a Tesla Model 3 interior layout with corresponding ambient lighting positions, integrated with a driving simulator and physical car seat for embodied experience.

\paragraph{Lighting Stimuli.} 
Lighting stimuli consisted of three conditions: blue ambient lighting (6500\,K color temperature), orange ambient lighting (2200\,K color temperature), and no ambient lighting control. We selected these colors based on research demonstrating their differential psychological effects \cite{Locken2017-ua}. Brightness levels were standardized at $200\pm5$\,lux measured at eye level, ensuring color tone rather than intensity served as the primary manipulated variable.

\paragraph{Olfactory Stimuli.} 
Scent stimuli consisted of three conditions: peppermint (\textit{Mentha piperita}), lavender (\textit{Lavandula angustifolia}), and no-scent control (water). We chose these essential oils based on documented contrasting effects on arousal and relaxation \cite{Herz2009-rn, Chen2024-hi}. Following pilot study results, essential oils were diluted to 1:50 ratio with neutral carrier oil and presented on absorbent pads within identical opaque jars to prevent visual cues. The no-scent control jar contained only carrier oil to ensure consistency. Figure~\ref{fig:experimental environment} illustrates the complete experimental apparatus and stimuli.

\subsection{Procedure}
\begin{figure}[htbp]
\centering
\includegraphics[width=.9\linewidth]{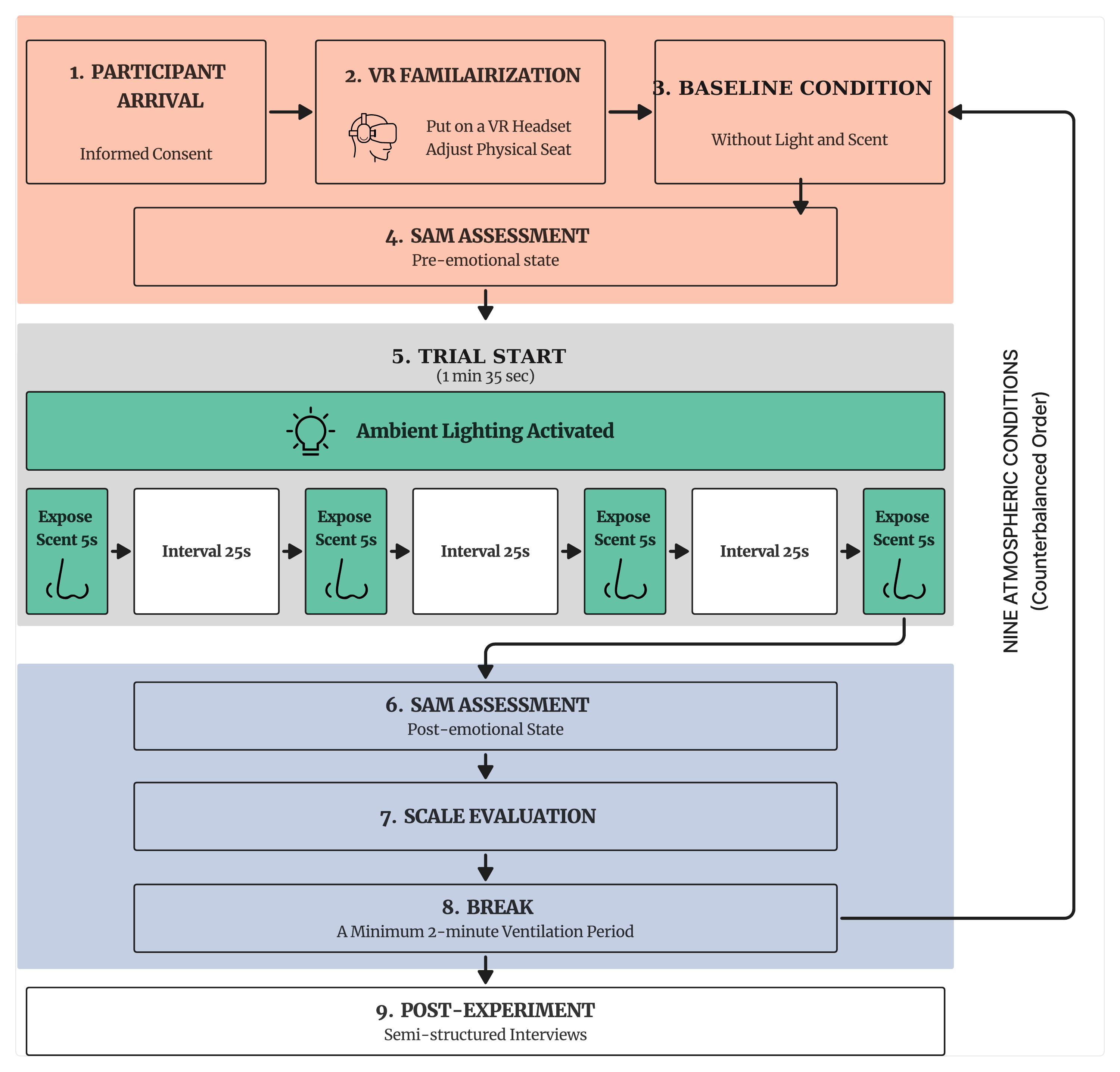}
\caption{Flowchart of the experimental procedure.}
\label{fig:experimental procedure}
\end{figure}

Figure~\ref{fig:experimental procedure} visualizes the complete experimental procedure. Upon arrival, participants provided written informed consent and were screened for conditions potentially interfering with the experiment (allergies, recent consumption of aromatic foods). Participants donned VR headsets and familiarized themselves with the simulated Tesla Model 3 interior from a first-person perspective, adjusting their physical car seats to align with virtual space. They received instructions that the vehicle operated in full self-driving mode to prime them for the passenger role. 

The experiment began with a 1-minute baseline condition without atmospheric stimuli. Participants then completed the Self-Assessment Manikin (SAM) to establish baseline pleasure and arousal levels \cite{Bradley1994-hd}. Following baseline assessment, participants experienced nine atmospheric conditions in counterbalanced order. Each trial lasted 1 minute 35 seconds. At trial start, assigned ambient lighting activated and remained on throughout. Simultaneously, the experimenter administered corresponding olfactory stimulus. Based on pilot study results, the scent jar was held approximately 4\,cm from participants' noses using a calibrated, non-intrusive physical guide affixed to the VR headset, ensuring precise and repeatable positioning. To minimize disruption to immersive experience, participants were not instructed to actively sniff. Instead, the scent source was held in calibrated position for initial 5-second passive olfactory exposure. To prevent olfactory adaptation \cite{Khan2007-lb, Velasco2014-mq}, the jar was removed and subsequently re-presented for three additional 5-second exposures at 25-second intervals, totaling four scent exposures per trial \cite{Ohtsu2009-ft}. Immediately after each trial, participants completed a second SAM assessment capturing emotional state changes, followed by questionnaires measuring five additional dimensions. A 2-minute rest period involving neutral cognitive tasks occurred between trials, with a minimum 2-minute ventilation period between scent conditions to minimize carryover effects \cite{Otto2021-jv}. 

After completing all nine conditions, participants engaged in brief semi-structured interviews about their experiences.

\subsection{Measurement Framework}
To empirically investigate our research questions and proposed S-O-R processes, we developed a comprehensive measurement framework. Construct selection was informed directly by our literature review, operationalizing key psychological mechanisms discussed in Section~2.3 and addressing research gaps identified in Section~2.4. Following S-O-R framework structure, we systematically categorized measures to capture each component of the environmental processing sequence.

\paragraph{Environmental Stimuli (S)}
The exogenous variables were the experimentally manipulated environmental conditions:
\begin{itemize}
  \item Lighting: 3 levels (blue, orange, none)
  \item Scent: 3 levels (mint, lavender, none)
\end{itemize}

\paragraph{Mediating Internal States (O)}
We measured several dimensions representing the organism's internal processing. Based on the psychological mechanisms identified in the literature, our measures were designed to capture a spectrum of potential processing stages, without pre-supposing a strict sequential order. These include:
\begin{itemize}
  \item \emph{Sensory Appraisal:} Capturing immediate hedonic responses to discrete, unimodal stimuli, measured by Sensory Preference for light and scent.
  \item \emph{Cross-Modal Integration:} Capturing the integrated hedonic response to the multimodal sensory blend, measured by Sensory Preference for combination.
  \item \emph{Higher-Order Environmental Appraisal:} Capturing more complex, cognitive evaluations of the environment, measured by Sensory Congruency and Perceived Atmosphere.
  \item \emph{Affective Response:} Capturing emotional impact as a parallel outcome of the appraisal process, measured by Emotional Change (Pleasure and Arousal scores from SAM).
\end{itemize}

\paragraph{Experiential Outcomes (R)}
The endogenous variables representing the final passenger responses were:
\begin{itemize}
  \item \emph{Holistic Journey Evaluation:} Measured by Overall Experience.
  \item \emph{Technology Acceptance:} Measured by Trust in the autonomous system.
\end{itemize}

Table~\ref{tab:measurement-framework} provides a comprehensive summary of this measurement framework, presenting each construct's theoretical definition, the specific measurement items used, its theoretical foundation, and its role within the S-O-R processing chain.

\begin{table}[!htbp]
\centering
\caption{Measurement Framework Following the S-O-R Processing Cascade}
\label{tab:measurement-framework}
\renewcommand{\arraystretch}{1.3}
\resizebox{\textwidth}{!}{%
\begin{tabular}{p{2.8cm} p{3.8cm} p{5.5cm} p{3.5cm} p{3.5cm}}
\hline
\textbf{Construct} & \textbf{Theoretical Definition} & \textbf{Measurement Items} & \textbf{Theoretical Foundation} & \textbf{Role in S-O-R Model} \\
\hline
\multicolumn{5}{l}{\textit{\textbf{O (Organism): Mediating Internal States}}} \\
\hline
\textbf{Sensory Appraisal} & Direct hedonic response to individual, unimodal sensory stimuli. & 
1. "I like the lighting in the cabin environment." Sensory preference (light) \newline 
2. "I like the scent in the cabin environment." Sensory preference (scent) & 
Spence et al. (2014); Reysen \& Hackett (2017) & 
Initial, unimodal perceptual processing. \\
\hline
\textbf{Cross-Modal Hedonic Integration} & Integrated hedonic response to the combined multisensory stimuli. & 
3. "I like the lighting and scent combination in the cabin environment." Sensory preference (combination) & 
Spence et al. (2014) & 
Integrated multimodal hedonic judgment. \\
\hline
\textbf{Higher-Order Environmental Appraisal} & Deliberative, cognitive evaluations of the environment's quality and coherence. & & & Complex cognitive evaluation. \\
\textit{--- Sensory Congruency} & \textit{Perceived harmony and compatibility between sensory modalities.} & 
1. "The lighting and scent in this cabin environment are well-matched." \newline 
2. "I feel that the lighting and scent in this cabin environment do not go well together." (reverse-scored) & 
Spangenberg et al. (2005); Demoulin (2011) & 
\textit{Complex cognitive evaluation of environmental coherence.} \\
\textit{--- Perceived Atmosphere} & \textit{Holistic evaluation of the emergent spatial experience created by integrated stimuli.} & 
1. "The overall atmosphere created by this cabin environment is pleasant." \newline 
2. "The overall atmosphere of this cabin environment is comfortable." \newline 
3. "The overall atmosphere of this cabin environment is appealing." & 
Bitner (1992); Rayburn \& Voss (2013) & 
\textit{Holistic cognitive evaluation of the environmental gestalt.} \\
\hline
\textbf{Affective Response (Emotion Change)} & Pre-post changes in emotional state across pleasure and arousal. & 
Self-Assessment Manikin (SAM; Bradley \& Lang, 1994) pictorial scale, with change scores calculated between baseline and post-exposure measurements. & 
Russell \& Mehrabian (1974); Bradley \& Lang (1994) & 
Emotional outcome of the appraisal process. \\
\hline
\multicolumn{5}{l}{\textit{\textbf{R (Response): Experiential Outcomes}}} \\
\hline
\textbf{Trust} & Confidence in and comfort with the autonomous driving system as influenced by environmental factors. & 
1. "I trust the autonomous driving system in this cabin environment." \newline 
2. "I feel skeptical about the autonomous driving system in this cabin environment." (reverse-scored) & 
Lee \& See (2004); Jian et al. (2000) & 
Key outcome variable (Automation Acceptance). \\
\hline
\textbf{Overall Experience} & Holistic evaluation of the quality of the autonomous journey experience as influenced by atmospheric elements. & 
1. "The overall travel experience in this cabin environment is satisfying." & 
Kolarova \& Cherchi (2021) & 
Primary outcome variable (Journey Evaluation). \\
\hline
\end{tabular}}
\begin{flushleft}
\footnotesize All dimensions except emotional changes were measured using 7-point Likert scales (1: strongly disagree to 7: strongly agree). The emotional change measures used the 9-point Self-Assessment Manikin pictorial scale, with change scores calculated by subtracting baseline from post-exposure ratings.
\end{flushleft}
\end{table}

\subsection{Data-Analysis Strategy}
We analyzed a 3 (Light: None/Blue/Orange) $\times$ 3 (Scent: None/Lavender/Mint) within-subjects design with 24 participants (216 observations). To maintain robust inference given small sample size, we adopted a sequential-evidence strategy:

\begin{enumerate}
  \item Document internal consistency of multi-item composites (Cronbach's $\alpha$, McDonald's $\omega$, corrected item-total correlations) using both raw and within-person centered scores
  \item Test direct experimental effects with repeated-measures ANOVA (R packages \texttt{afex}/\texttt{emmeans}), applying Greenhouse-Geisser corrections when $\epsilon < 1$ and Bonferroni-adjusted pairwise comparisons
  \item Explore within-person S-O-R pathways using mixed-effects regressions (random intercepts for participants; person-mean centering) and participant-cluster bootstrap to quantify total indirect effects from stimuli to outcomes
\end{enumerate}

Because $N=24$ proves underpowered for confirmatory latent modeling, all confirmatory inference relies on approaches (ii) and (iii), treating pathway exploration as hypothesis-generating rather than confirmatory analysis.

\paragraph{Construct Handling at Small $N$.}
Perceived Atmosphere, Sensory Congruency, and Trust were analyzed as composites rather than latent variables in the main analytical pipeline. We report $\alpha$, $\omega$, and corrected item-total correlations (CITC) on both raw and within-centered scores. For 2-item composites (Sensory Congruency, Trust), $\alpha = \omega$ by identity; interpretation follows composite rather than latent variable perspective.

\paragraph{Direct Effects (RQ1).}
For each dependent variable, we fit 3$\times$3 repeated-measures ANOVA in R/\texttt{afex}, reporting omnibus $F$, $p$, and partial $\eta^2_p$. We applied Greenhouse-Geisser corrections where needed and obtained Bonferroni pairwise contrasts via R/\texttt{emmeans}. This approach directly addresses atmospheric effectiveness in confined mobile environments.

\paragraph{Sequential Processing Pathway Exploration (RQ2).}
To identify psychological processes mediating environmental effects, we employed exploratory pathway analysis. The analytical framework development proceeded iteratively: initial correlational analyses and experimental results guided decomposition of measured constructs into potential processing stages, which were then formally tested through pathway modeling. This data-informed approach enables identification of the most parsimonious sequential structure capturing atmospheric mechanisms.

Predictors were person-mean centered; stimuli were treatment-coded (reference = "None"). Under strict within-person centering in balanced 9-cell design, random-intercept variance collapses (singular fit), so estimates coincide with participant fixed-effects; inference relies on fixed-effects/cluster-robust logic. For stimulus to outcome total indirect effects, we employed participant-cluster bootstrap ($B=5{,}000$) for robust confidence intervals.

\paragraph{Affect Scale Harmonization.}
Pleasure and Arousal represent SAM deltas (1-9 scale), while other items use 7-point Likert scales. We tested robustness by adding within-person $z$-scored Pleasure and Arousal to outcome models. Incremental explanatory power proved minimal ($\Delta R^2_m \approx -0.002$ for both outcomes), supporting the parsimonious evaluative route via Perceived Atmosphere and Trust.

\paragraph{Exploratory Structural Equation Model (SEM).}
To triangulate sequential-regression findings with integrated representation of the S-O-R cascade, we estimated within-person SEM in the Supplement (Section S2). The SEM applies person-mean centering, MLR/FIML estimation, and participant-clustered standard errors. Given small sample size, SEM results are interpreted descriptively as hypothesis-generating evidence rather than confirmatory inference.

\paragraph{Software.}
All analyses were conducted in R (rm-ANOVA: \texttt{afex}; post-hoc: \texttt{emmeans}; mixed models: \texttt{lme4}/\texttt{lmerTest}; bootstrap: custom participant-cluster resampling).

\section{Findings} 
Analysis proceeded in three stages. First, we examined measurement properties and descriptive patterns (Section 4.1). Second, we tested direct atmospheric effects using repeated-measures ANOVA, addressing RQ1 (Section 4.2). Third, we modeled psychological pathways through sequential regression and structural equation modeling, addressing RQ2 (Section 4.3). Qualitative interviews provided additional 
validation (Section 4.4).

\subsection{Measurement and Descriptive Patterns}
Descriptive statistics and correlational analyses used raw (uncentered) scores to preserve interpretability on original measurement scales. Subsequent analyses (Sections 4.2-4.3) employed within-person centering to isolate trial-level variance, consistent with our focus on atmospheric effects within individuals.

\subsubsection{Internal Consistency and Item Diagnostics}
Multi-item constructs were analyzed as unit-weighted composites given $N=24$. Perceived Atmosphere showed excellent reliability (raw $\alpha = .92$, $\omega = .95$; within-centered $\alpha = .918$, $\omega = .948$). For two-item composites, Sensory Congruency was acceptable (raw/within $\alpha = \omega \approx .57/.62$), while Trust improved substantially after within-person centering (raw $\alpha = .38$ $\rightarrow$ within $\alpha = .717$). Corrected item-total correlations were high for Perceived Atmosphere (.81-.86), moderate for Sensory Congruency (~.45), and improved for Trust after centering (raw ~.23 $\rightarrow$ within ~.56). Tables~\ref{tab:alpha} and \ref{tab:citc} provide full details.

\begin{table}[htbp]
\centering
\caption{Internal consistency ($\alpha/\omega$) for multi-item constructs (raw and within-centered)}
\label{tab:alpha}
\begin{tabular}{lllll}
\hline
Spec & Scale & k & Alpha & Omega \\
\hline
Raw             & Perceived Atmosphere & 3 & 0.92  & 0.95  \\
Raw             & Sensory Congruency         & 2 & 0.573 & 0.573 \\
Raw             & Trust      & 2 & 0.378 & 0.378 \\
Within-centered & Perceived Atmosphere & 3 & 0.918 & 0.948 \\
Within-centered & Sensory Congruency         & 2 & 0.62  & 0.62  \\
Within-centered & Trust      & 2 & 0.717 & 0.717 \\
\hline
\end{tabular}
\end{table}

\begin{table}[htbp]
\centering
\caption{Corrected item--total correlations (CITC)}
\label{tab:citc}
\begin{tabular}{lll}
\hline
Scale & Item & CITC \\
\hline
\multicolumn{3}{l}{(A) Raw items} \\
Perceived Atmosphere & perceived atmosphere\_appeal      & 0.82  \\
Perceived Atmosphere & perceived atmosphere\_comfort     & 0.849 \\
Perceived Atmosphere & perceived atmosphere\_pleasure    & 0.847 \\
Sensory Congruency         & Sensory Congruency\_coord              & 0.401 \\
Sensory Congruency         & Sensory Congruency\_intrusive\_r       & 0.401 \\
Trust      & trust\_pos             & 0.233 \\
Trust      & trust\_neg\_r          & 0.233 \\
\hline
\multicolumn{3}{l}{(B) Within-centered items} \\
Perceived Atmosphere & perceived atmosphere\_appeal\_cw   & 0.809 \\
Perceived Atmosphere & perceived atmosphere\_comfort\_cw  & 0.856 \\
Perceived Atmosphere & perceived atmosphere\_pleasure\_cw & 0.838 \\
Sensory Congruency         & Sensory Congruency\_coord\_cw           & 0.449 \\
Sensory Congruency         & Sensory Congruency\_intrusive\_r\_cw    & 0.449 \\
Trust      & trust\_pos\_cw          & 0.558 \\
Trust      & trust\_neg\_r\_cw       & 0.558 \\
\hline
\end{tabular}
\begin{flushleft}
\footnotesize Note. ``\_cw'' denotes person-mean centered indicators.
\end{flushleft}
\end{table}

\subsubsection{Descriptive Statistics and Distributional Notes}

\begin{table}[htbp]
\centering
\caption{Descriptive statistics for selected variables ($N=216$ observations)}
\label{tab:descriptives}
\begin{tabular}{lrrrrrr}
\hline
Variable & Min & Max & M & SD & Skewness & Kurtosis \\
\hline
Sensory preference\_light           & 1  & 7  & 4.41 & 1.54 & -0.21 & -0.48 \\
Sensory preference\_scent           & 1  & 7  & 4.58 & 1.46 & -0.34 & -0.12 \\
Sensory preference\_combination     & 1  & 7  & 4.34 & 1.54 & -0.18 & -0.52 \\
Sensory Congruency                  & 2  & 14 & 8.65 & 2.55 & -0.02 & -0.31 \\
Perceived Atmosphere          & 3  & 21 & 12.65 & 4.18 & -0.84 & 0.89 \\
Trust               & 2  & 14 & 8.55 & 2.27 & -0.12 & -0.18 \\
Overall Experience  & 1  & 7  & 4.06 & 1.56 & -0.09 & -0.41 \\
Pleasure Change     & -6 & 4  & 0.37 & 1.59 & -0.25 & 0.33 \\
Arousal Change      & -4 & 8  & 1.04 & 1.64 & 1.23  & 2.11 \\
\hline
\end{tabular}
\end{table}

Table~\ref{tab:descriptives} summarizes all measures (216 observations, 24 participants $\times$ 9 conditions). Preference indices showed substantial dispersion (Sensory preference light: $M=4.41$, $SD=1.54$; scent: $M=4.58$, $SD=1.46$; combination: $M=4.34$, $SD=1.54$), indicating trial-level sensitivity to manipulations. Pleasure change centered modestly above zero ($M=0.37$, $SD=1.59$); arousal change was 
positively skewed ($M=1.04$, $SD=1.64$, skew=1.23). Perceived Atmosphere spanned a wide range ($SD=4.18$) with mild negative skew (-0.84). These patterns document substantial within-person variation motivating mixed-effects models using person-mean centering.

\subsubsection{Correlational Structure}
\begin{figure}[htbp]
\centering
\includegraphics[width=.9\linewidth]{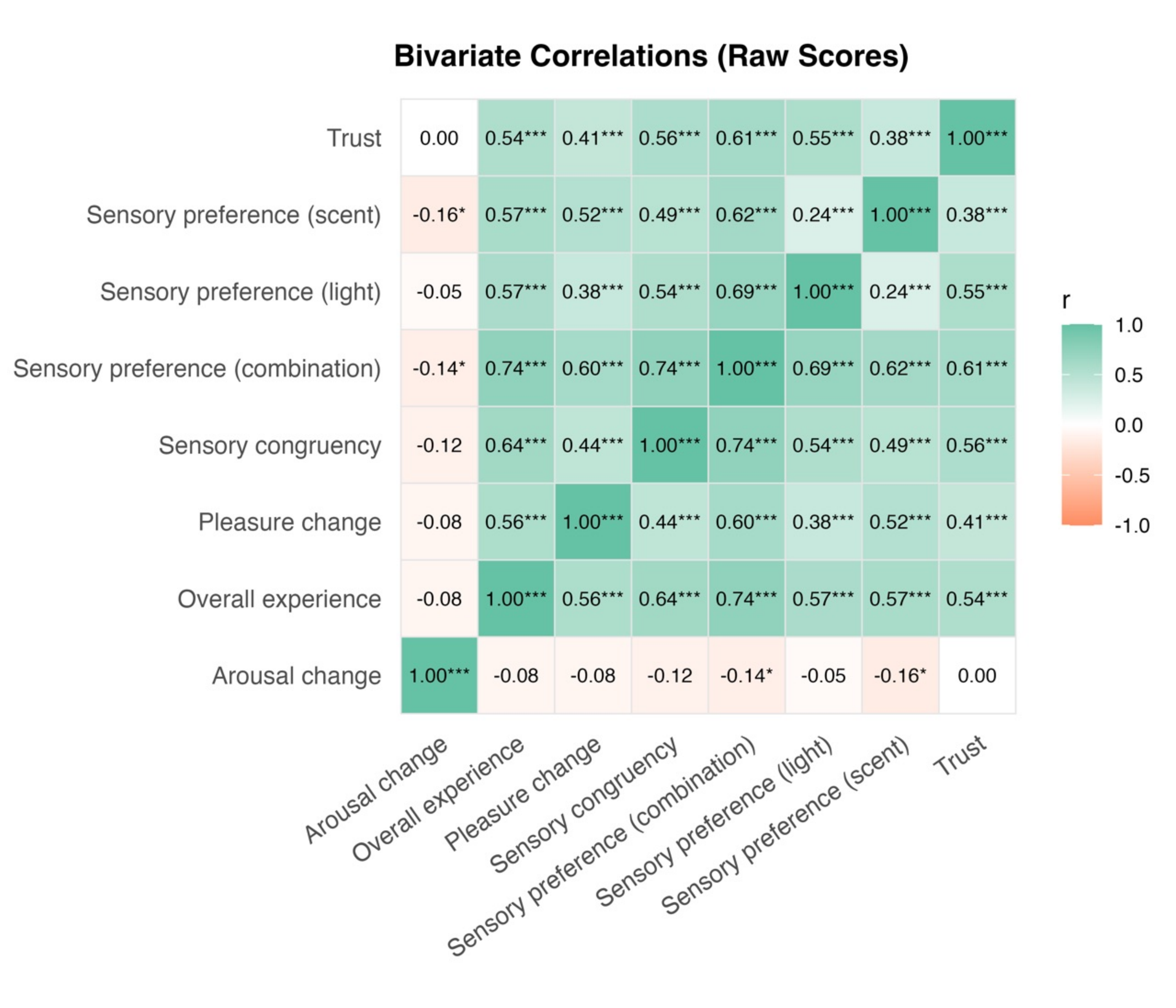}
\caption{Bivariate correlations between all measured variables}
\label{fig:correlation}
\end{figure}

Figure~\ref{fig:correlation} displays bivariate correlations. Sensory preference light and scent correlated modestly ($r=.24$, $p<.01$), suggesting independent initial appraisal; both strongly predicted their integration (light$\rightarrow$combination: $r=.69$; scent$\rightarrow$combination: $r=.62$; both $p<.001$), consistent with additive cross-modal integration.

Perceived Atmosphere occupied a central position, correlating highly  with Sensory preference combination ($r=.84$, $p<.001$) and Overall  Experience ($r=.80$, $p<.01$), and substantially with Sensory  preference scent ($r=.73$, $p<.001$) and Pleasure Change ($r=.66$,  $p<.001$). Despite high correlation, Sensory preference combination  and Perceived Atmosphere remain conceptually distinct: the former  reflects immediate hedonic appraisal of the sensory blend, the latter  indexes higher-order environmental gestalt.  Notably, Sensory preference scent showed tighter links with Perceived  Atmosphere ($r=.73$) than Sensory preference light (~.60), and  comparable links with Overall Experience ($r=.57$, $p<.01$). This  olfactory centrality diverges from typical visual dominance in  transport contexts. Trust related moderately to Perceived Atmosphere  ($r=.58$, $p<.01$) and Sensory preference combination ($r=.61$,  $p<.01$), foreshadowing pathway results where Perceived Atmosphere  and Sensory Congruency inform Trust.

\subsection{Direct Effects of Light and Scent (RQ1)}
\begin{figure}[htbp]
\centering
\includegraphics[width=\linewidth]{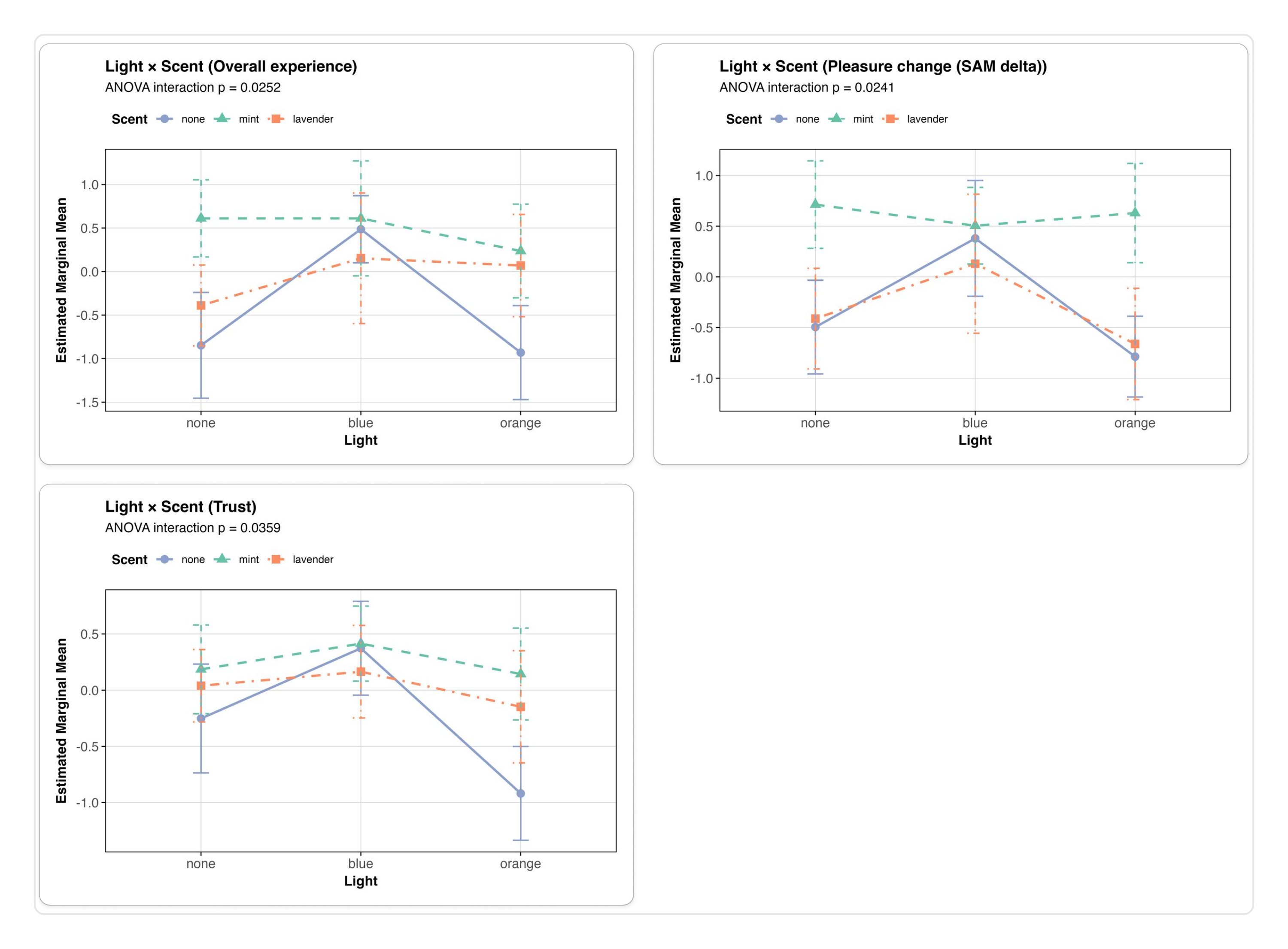}
\caption{Significant Light $\times$ Scent interaction effects}
\label{fig:interaction}
\end{figure}

\begin{figure}[htbp]
\centering
\includegraphics[width=.8\linewidth]{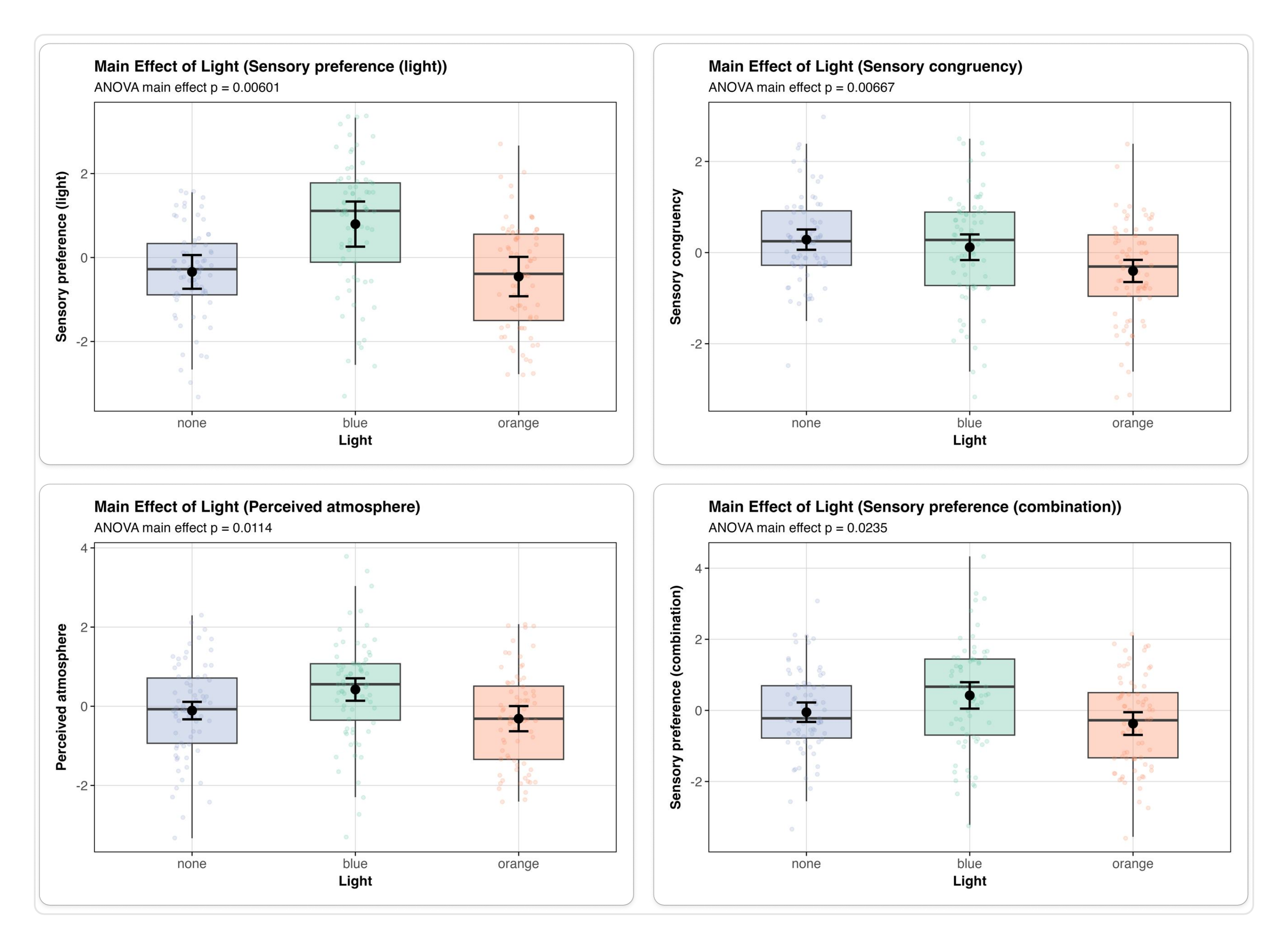}
\caption{Significant main effects of Lighting}
\label{fig:Lighting}
\end{figure}

\begin{figure}[htbp]
\centering
\includegraphics[width=.8\linewidth]{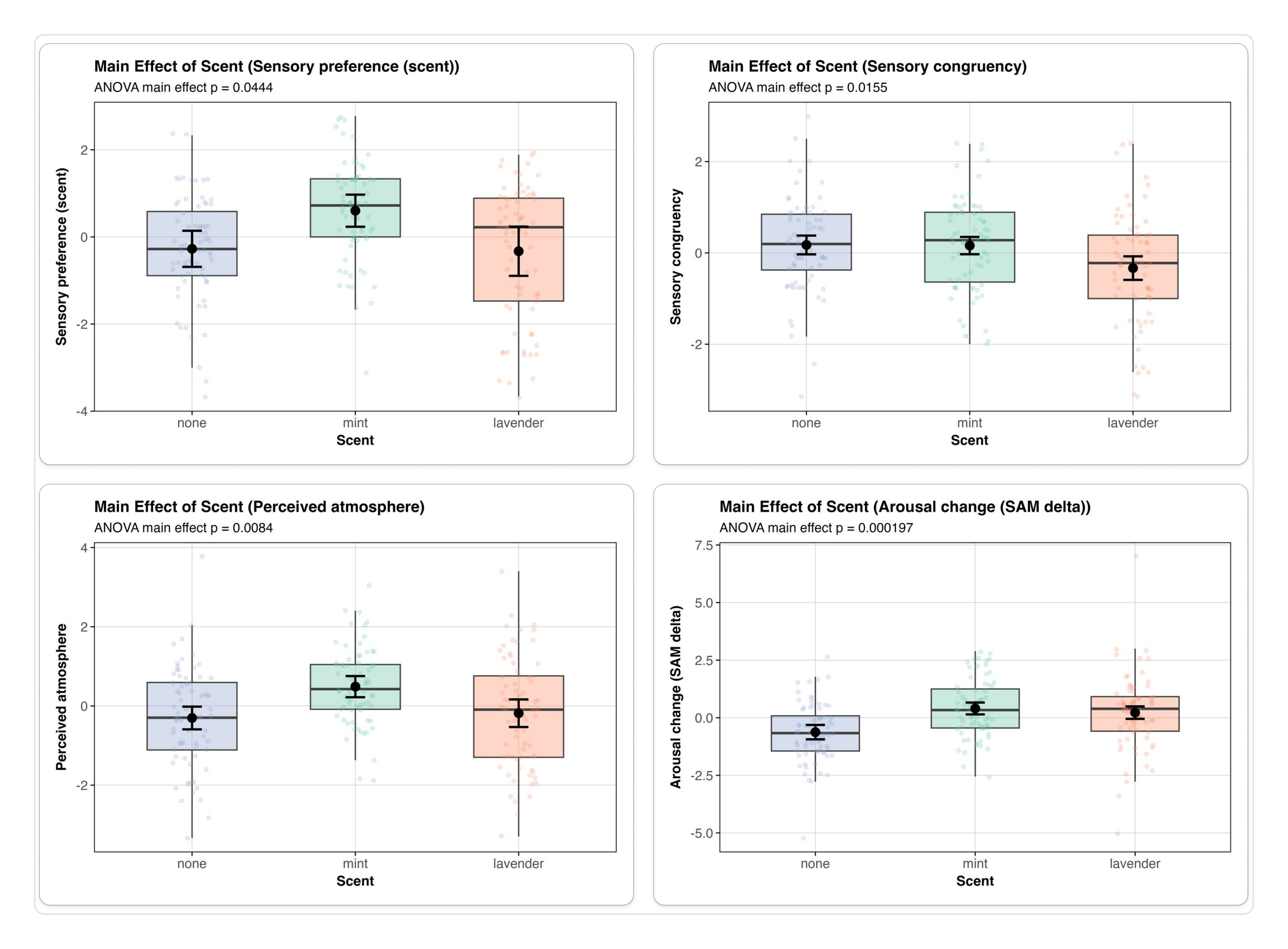}
\caption{Significant main effects of Scent}
\label{fig:Scent}
\end{figure}

We tested whether lighting and scent influenced preferences,  integration, appraisals, affect, and experiential outcomes using  3$\times$3 repeated-measures ANOVAs with Greenhouse-Geisser corrections  where sphericity was violated ($\epsilon<1$) and Bonferroni post-hoc  tests. Table~\ref{tab:anova} summarizes omnibus tests. Interaction  plots appear in Figure~\ref{fig:interaction} (Pleasure, Overall  Experience, Trust). Main-effect plots appear in Figures~\ref{fig:Lighting}  and \ref{fig:Scent} when main effects were significant without  interactions (Sensory preferences, Perceived Atmosphere, Sensory Congruency, Arousal).

\paragraph{Sensory Appraisal (Unimodal).}
Lighting preference showed a robust Light main effect, $F=6.11$,  $p=.006$, $\eta^2_p=.21$. Blue exceeded Orange and None (both  $p \leq .031$); Orange did not differ from None. Scent had no effect;  no interaction emerged, consistent with modality-specific visual evaluation.

Scent preference showed a Scent main effect, $F=3.72$, $p=.044$,  $\eta^2_p=.14$. Mint exceeded None ($p=.0088$); Lavender did not  differ from other levels. Lighting had no effect; no interaction  emerged, paralleling the lighting preference result and reinforcing channel specificity in initial appraisal.

\paragraph{Cross-Modal Integration.}
Combined preference showed a Light main effect, $F=4.31$, $p=.023$,  $\eta^2_p=.16$, and marginal Scent effect, $F=3.10$, $p=.068$,  $\eta^2_p=.12$. Blue tended to exceed Orange (Bonferroni $p=.051$).  No interaction emerged, supporting additive integration rather than synergistic amplification.

\paragraph{Higher-Order Environmental Appraisal.}
Perceived Atmosphere showed both Light and Scent effects (Light:  $F=5.37$, $p=.011$, $\eta^2_p=.19$; Scent: $F=5.66$, $p=.008$,  $\eta^2_p=.20$). Blue exceeded Orange ($p=.037$); Mint exceeded None  ($p=.0028$). The interaction was marginal ($F=2.31$, $p=.087$).  Olfactory input matched visual input in magnitude for shaping environmental gestalt.

Sensory Congruency showed both Light and Scent effects (Light:  $F=5.86$, $p=.007$, $\eta^2_p=.20$; Scent: $F=4.93$, $p=.015$,  $\eta^2_p=.18$), without interaction. Orange was perceived as less coherent than None (Bonferroni $p=.0026$).

\paragraph{Affective Responses.}
Pleasure was dominated by Scent, $F=8.24$, $p=.002$, $\eta^2_p=.26$:  Mint exceeded None ($p<.001$) and Lavender ($p=.0036$). A Light$\times$Scent  interaction emerged ($F=3.17$, $p=.024$), shown in Figure~\ref{fig:interaction}: mint's  pleasure lift was pronounced under Orange and None but attenuated under Blue.

Arousal showed the largest effect in the study: Scent $F=10.84$,  $p<.001$, $\eta^2_p=.32$. Mint exceeded None; Lavender exceeded None (both $p \leq .009$). Lighting was marginal; no interaction emerged,  implying olfactory-driven activation.

\paragraph{Experiential Outcomes.}
Overall Experience improved with Scent, $F=6.35$, $p=.005$,  $\eta^2_p=.22$, driven by Mint>None ($p=.001$). A Light$\times$Scent  interaction emerged ($F=3.20$, $p=.025$; Figure~\ref{fig:interaction}): scent benefits  were largest under Orange or None, whereas Blue reduced reliance on scent.

Trust showed both Light and Scent effects (Light: $F=3.84$, $p=.032$,  $\eta^2_p=.143$; Scent: $F=4.34$, $p=.020$, $\eta^2_p=.159$), with  a Light$\times$Scent interaction ($F=2.95$, $p=.036$; Figure~\ref{fig:interaction}). Under  Orange, Mint and Lavender exceeded None (mint-none $p=.0044$;  lavender-none $p=.0112$); under Blue, scents did not differ. Within  No-scent, Blue exceeded Orange (blue-orange $p=.0046$). Orange  lighting depressed Trust unless compensated by scent; blue lighting  attenuated reliance on olfactory support.

\begin{table}[htbp]
\centering
\caption{rm-ANOVA omnibus tests by dependent variable (Greenhouse--Geisser where applicable)}
\label{tab:anova}
\begin{tabular}{lllll}
\hline
DV & Effect & $F$ & $p$ & $\eta^2_p$ \\
\hline
Sensory preference\_light (cw)       & Light           & 6.114  & 0.006 & 0.21  \\
                     & Scent           & 0.421  & 0.629 & 0.018 \\
                     & Light $\times$ Scent & 1.113  & 0.351 & 0.046 \\
Sensory preference\_scent (cw)       & Light           & 0.086  & 0.904 & 0.004 \\
                     & Scent           & 3.717  & 0.044 & 0.139 \\
                     & Light $\times$ Scent & 0.419  & 0.757 & 0.018 \\
Sensory preference\_combination (cw) & Light           & 4.313  & 0.023 & 0.158 \\
                     & Scent           & 3.095  & 0.068 & 0.119 \\
                     & Light $\times$ Scent & 1.137  & 0.336 & 0.047 \\
Perceived Atmosphere (cw)      & Light           & 5.373  & 0.011 & 0.189 \\
                     & Scent           & 5.657  & 0.008 & 0.197 \\
                     & Light $\times$ Scent & 2.313  & 0.087 & 0.091 \\
Sensory Congruency (cw) & Light        & 5.864  & 0.007 & 0.203 \\
                     & Scent           & 4.932  & 0.015 & 0.177 \\
                     & Light $\times$ Scent & 1.262  & 0.294 & 0.052 \\
Pleasure change (cw) & Light           & 3.471  & 0.057 & 0.131 \\
                     & Scent           & 8.235  & 0.002 & 0.264 \\
                     & Light $\times$ Scent & 3.167  & 0.024 & 0.121 \\
Arousal change (cw)  & Light           & 3.292  & 0.060 & 0.125 \\
                     & Scent           & 10.844 & $<.001$ & 0.320 \\
                     & Light $\times$ Scent & 0.890  & 0.460 & 0.037 \\
Overall Experience (cw) & Light        & 3.354  & 0.056 & 0.127 \\
                     & Scent           & 6.351  & 0.005 & 0.216 \\
                     & Light $\times$ Scent & 3.201  & 0.025 & 0.122 \\
Trust (cw)           & Light           & 3.840  & 0.032 & 0.143 \\
                     & Scent           & 4.340  & 0.020 & 0.159 \\
                     & Light $\times$ Scent & 2.950  & 0.036 & 0.114 \\
\hline
\end{tabular}
\begin{flushleft}
\footnotesize Note. Full Bonferroni matrices are provided in Supplement~\ref{supp:S1}.
\end{flushleft}
\end{table}

\subsection{Modeling the Within-Person Pathway (RQ2)}
While ANOVAs confirmed atmospheric influences on passenger experience, they did not reveal underlying mechanisms. Correlational patterns (Section 4.1.3) suggested sequential processing: sensory-specific appraisals (light, scent preferences) correlated modestly ($r=.24$), implying independent channels, yet both strongly predicted their integration (combination preference), which in turn strongly predicted holistic gestalt (Perceived Atmosphere). Experimental results (Section 4.2) reinforced this structure: multisensory interactions were absent at basic preference levels but emerged for higher-order outcomes (Trust, Overall Experience), implying independent initial processing followed by integration at later evaluative stages.

\begin{figure}[htbp]
\centering
\includegraphics[width=.9\linewidth]{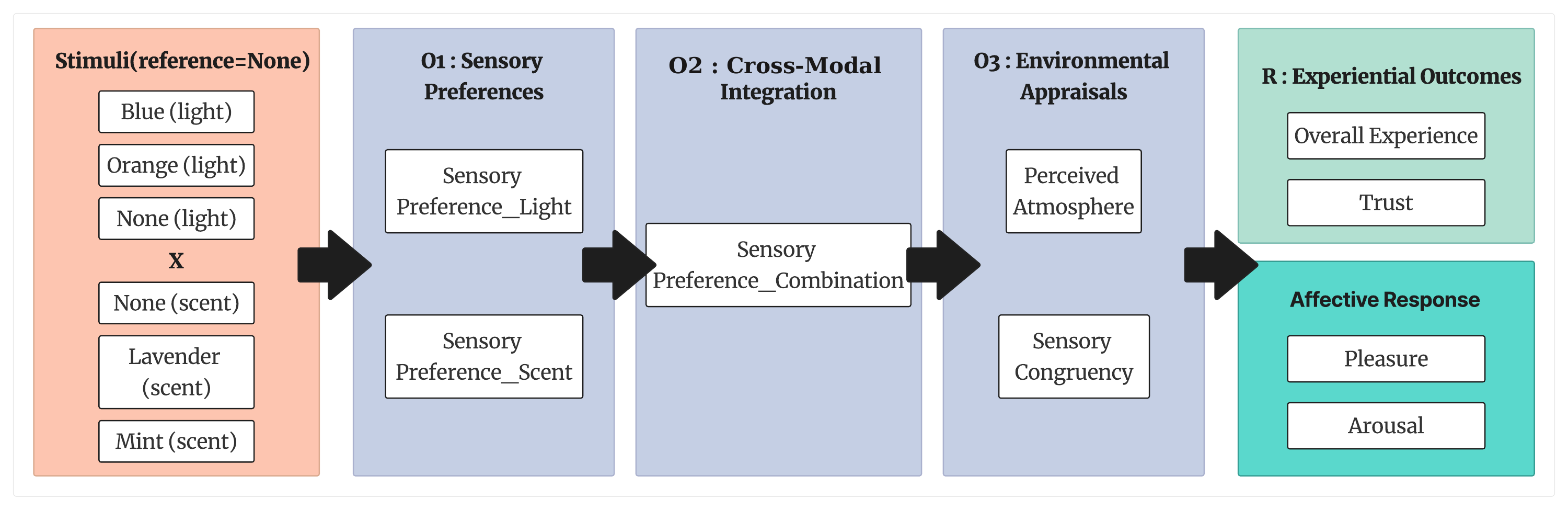}
\caption{The hypothesized S-O-R conceptual schematic}
\label{fig:sor_concept}
\end{figure}

Based on this convergent evidence, we tested a sequential S-O-R pathway model. Figure~\ref{fig:sor_concept} presents the conceptual schematic formally tested through sequential regressions.

\noindent
\textit{Notes for Fig.~\ref{fig:sor_concept}.} Solid arrows denote supported links in Table~\ref{tab:serial_regs}; dashed arrows denote tested but not required links (e.g., Orange/Lavender to single-channel preferences; Trust $\rightarrow$ Overall Experience; and Perceived Atmosphere $\rightarrow$ Pleasure/Arousal shown as exploratory). Variables are person-mean centered; stimuli are treatment-coded (baseline = ``None''). Numerical estimates are reported in Tables~\ref{tab:serial_regs}--\ref{tab:trust_decomp} and supplementary \ref{supp:S5}.

\subsubsection{Serial regressions by stage}

Table~\ref{tab:serial_regs} presents estimates. At O1 (initial appraisal), Blue$\rightarrow$Sensory preference light ($b=1.139$, $p<.001$) and Mint$\rightarrow$Sensory preference scent ($b=0.875$, $p<.001$), with opposite cross-effects absent, replicating modality-specific patterns from Section 4.2. At O2 (integration), Sensory preference combination aggregated channel preferences additively (light$\rightarrow$combination $b=0.567$; scent$\rightarrow$combination $b=0.474$; both $p<.001$). At O3 (higher-order appraisal), Sensory preference combination predicted both Sensory Congruency ($b=0.607$, $p<.001$) and Perceived Atmosphere ($b=0.422$, $p<.001$), while Sensory preference light ($b=0.195$, $p<.001$) and scent ($b=0.366$, $p<.001$) contributed uniquely to Perceived Atmosphere. In the response layer, Perceived Atmosphere and Sensory Congruency jointly predicted Trust ($b=0.366$ and $b=0.198$, respectively; both $p \leq .003$), and Perceived Atmosphere dominated Overall Experience prediction ($b=0.780$, $p<.001$) with Trust adding smaller parallel contribution ($b=0.230$, $p=.001$). The resulting architecture shows preference$\rightarrow$integration$\rightarrow$appraisal hub driving both Trust and Overall Experience.

\begin{table}[htbp]
\centering
\caption{Serial within-person regressions by stage (random intercept for participant)}
\label{tab:serial_regs}
\begin{tabularx}{\textwidth}{l X c c c c}
\toprule
Stage & Predictor $\rightarrow$ Outcome & Estimate & SE & df & $p$ \\
\midrule
O1 (S$\rightarrow$Channel) 
 & \texttt{l\_blue} $\rightarrow$ \texttt{Sensory preference\_light\_cw} & 1.139 & 0.219 & 213 & $<.001$ \\
 & \texttt{l\_orange} $\rightarrow$ \texttt{Sensory preference\_light\_cw} & $-0.111$ & 0.219 & 213 & .613 \\
O1 (S$\rightarrow$Channel) 
 & \texttt{s\_mint} $\rightarrow$ \texttt{Sensory preference\_scent\_cw} & 0.875 & 0.217 & 213 & $<.001$ \\
 & \texttt{s\_lav}  $\rightarrow$ \texttt{Sensory preference\_scent\_cw} & $-0.056$ & 0.217 & 213 & .798 \\
O2 (Integration) 
 & \texttt{Sensory preference\_light\_cw} $\rightarrow$ \texttt{Sensory preference\_comb\_cw} & 0.567 & 0.037 & 213 & $<.001$ \\
 & \texttt{Sensory preference\_scent\_cw} $\rightarrow$ \texttt{Sensory preference\_comb\_cw} & 0.474 & 0.039 & 213 & $<.001$ \\
O3 (Appraisal) 
 & \texttt{Sensory preference\_comb\_cw} $\rightarrow$ \texttt{Sensory Congruency\_cw} & 0.607 & 0.038 & 214 & $<.001$ \\
 & \texttt{Sensory preference\_comb\_cw} $\rightarrow$ \texttt{Perceived Atmosphere\_cw} & 0.422 & 0.050 & 212 & $<.001$ \\
 & \texttt{Sensory preference\_light\_cw} $\rightarrow$ \texttt{Perceived Atmosphere\_cw} & 0.195 & 0.039 & 212 & $<.001$ \\
 & \texttt{Sensory preference\_scent\_cw} $\rightarrow$ \texttt{Perceived Atmosphere\_cw} & 0.366 & 0.037 & 212 & $<.001$ \\
R (Trust) 
 & \texttt{Perceived Atmosphere\_cw} $\rightarrow$ \texttt{Trust\_cw} & 0.366 & 0.059 & 213 & $<.001$ \\
 & \texttt{Sensory Congruency\_cw} $\rightarrow$ \texttt{Trust\_cw} & 0.198 & 0.066 & 213 & .003 \\
R (Overall Experience) 
 & \texttt{Perceived Atmosphere\_cw} $\rightarrow$ \texttt{Overall Experience\_cw} & 0.780 & 0.057 & 213 & $<.001$ \\
 & \texttt{Trust\_cw} $\rightarrow$ \texttt{Overall Experience\_cw} & 0.230 & 0.070 & 213 & .001 \\
\bottomrule
\end{tabularx}
\begin{flushleft}
\footnotesize Notes. All variables are person-mean centered. Stimuli are treatment-coded (reference = ``None''). df by Satterthwaite. Random-intercept fits are singular under strict within-centering; results align with fixed-effects OLS.
\end{flushleft}
\end{table}

\subsubsection{Exploratory SEM as an integrated triangulation}

An exploratory within-person SEM (person-mean centered; MLR/FIML; participant-clustered standard errors) provided integrated assessment of the S-O-R cascade. Model fit was acceptable for exploratory specification (Model A: CFI=.924, TLI=.899, RMSEA=.093, SRMR=.049; robust CFI=.935, TLI=.913, robust RMSEA=.085). Nested comparison showed removing Trust$\rightarrow$Overall Experience path did not degrade fit (Satorra-Bentler $\Delta\chi^2(2)=2.79$, $p=.248$; $R^2$(Overall Experience)=.665 vs .660), indicating Trust adds only modest parallel contribution once Perceived Atmosphere is included. Standardized pattern replicated sequential regressions: Blue$\rightarrow$Sensory preference light=.378; Mint$\rightarrow$Sensory preference scent=.292; Sensory preference combination$\rightarrow$Perceived Atmosphere=.486; Perceived Atmosphere$\rightarrow$Overall Experience=.814. Full specification appears in Supplement~\ref{supp:S2}.

\subsubsection{Stimulus-to-Overall-Experience indirect effects (participant-cluster bootstrap)}

We quantified total within-subject indirect effects from stimuli to Overall Experience using participant fixed-effects OLS with participant-cluster bootstrap ($B=5{,}000$). Both Blue and Mint yielded reliable totals (Blue$\rightarrow$Overall Experience=$0.445$, $SE=0.154$, $p=.005$; Mint$\rightarrow$Overall Experience=$0.439$, $SE=0.150$, $p<.001$). Portions traveling exclusively through Trust were small (Blue via-Trust=$0.059$, $SE=0.038$, $p=.023$; Mint via-Trust=$0.053$, $SE=0.031$, $p=.019$), aligning with sequential regressions and SEM comparison. Tables~\ref{tab:indirect_total} and \ref{tab:trust_decomp} provide full decomposition.

\begin{table}[htbp]
\centering
\caption{Total indirect effects of stimuli on Overall Experience (cluster bootstrap)}
\label{tab:indirect_total}
\begin{tabular}{lrrr}
\hline
Path & Estimate & $SE_{\text{boot}}$ & $p_{\text{boot}}$ \\
\hline
Blue $\rightarrow$ Overall Experience (total indirect) & 0.445 & 0.154 & .005 \\
Mint $\rightarrow$ Overall Experience (total indirect) & 0.439 & 0.150 & $<.001$ \\
\hline
\end{tabular}
\begin{flushleft}
\footnotesize Notes. Participant-fixed effects with cluster bootstrap ($B=5{,}000$).
\end{flushleft}
\end{table}

\begin{table}[htbp]
\centering
\caption{Decomposition and total effects to Trust}
\label{tab:trust_decomp}
\begin{tabular}{lrrr}
\hline
Quantity & Estimate & $SE_{\text{boot}}$ & $p_{\text{boot}}$ \\
\hline
via Trust only (Blue $\rightarrow$ Overall Experience) & 0.059 & 0.038 & .023 \\
via Trust only (Mint $\rightarrow$ Overall Experience) & 0.053 & 0.031 & .019 \\
Blue $\rightarrow$ Trust (total)       & 0.259 & 0.100 & .005 \\
Mint $\rightarrow$ Trust (total)       & 0.231 & 0.085 & $<.001$ \\
\hline
\end{tabular}
\begin{flushleft}
\footnotesize Notes. ``via Trust only'' aggregates paths that reach Overall Experience exclusively through Trust; magnitudes are small relative to the total indirect, consistent with Table~\ref{tab:serial_regs} where Perceived Atmosphere dominates Overall Experience.
\end{flushleft}
\end{table}

\subsubsection{Affect as parallel responses}
As a manipulation check, Perceived Atmosphere predicted Pleasure ($b=0.753$, $SE=0.048$, $t(214)=15.7$, $p<.001$) and Arousal ($b=-0.246$, $SE=0.078$, $t(214)=-3.15$, $p=.0019$) within persons. However, adding $z$-scored Pleasure/Arousal to Overall Experience and Trust models did not improve explanatory power ($\Delta R^2_m \approx -0.002$ for both; joint Wald for affect terms in Trust model=$0$, $p=1.00$). Affect behaves as parallel responses to appraisals rather than additional determinants of Overall Experience or Trust.

\subsection{Qualitative Pattern Validation}
Semi-structured interviews ($n=24$, 20--30 minutes) using reflexive thematic analysis validated three core empirical patterns while providing mechanistic insights into the psychological processes underlying quantitative findings.

\subsubsection{Individual Variation in Sensory Processing}
Interview data revealed substantial individual differences in sensory prioritization strategies that help explain the systematic olfactory prominence observed quantitatively. Participants described context-dependent sensory dominance patterns:

\begin{quote}
``If these two [stimuli] are conflicting, I'm more affected by the scent... you can selectively choose not to look at it, but with scent, there's no way to avoid it.'' (P14)
\end{quote}

\begin{quote}
``I think visuals are more influential. The scent was intermittent, but the light distribution was broader.'' (P11)
\end{quote}

Seven participants explicitly reported that conflicting stimuli created distinct experiential states rather than one modality dominating:

\begin{quote}
``When these two things conflict, I feel more uncomfortable, rather than relaxed or tense.'' (P2)
\end{quote}

These accounts provide mechanistic insight into the compensatory dynamics identified quantitatively, suggesting passengers engage in active sensory management strategies.

\subsubsection{Environmental Coherence and System Assessment}
Participants consistently connected environmental design quality to technology competence evaluations, validating the dual pathways to trust identified statistically. Environmental coherence served as a heuristic for system reliability:

\begin{quote}
``If the colors and scents feel mismatched, it makes me question whether the system is coherent... it affects my confidence in the technology.'' (P18)
\end{quote}

Several participants described technological associations:

\begin{quote}
``The blue lighting feels technological and competent---it aligns with what I expect from an advanced system.'' (P4)
\end{quote}

These observations confirm that the Sensory Congruency$\rightarrow$Trust pathway ($\beta = 0.169$, $p = .037$) operates through symbolic evaluation of system sophistication.

\subsubsection{Context-Dependent Atmospheric Preferences}
All participants described fundamental shifts in environmental preferences between manual and autonomous driving contexts, supporting our focus on passenger-specific design principles. Representative responses included:

\begin{quote}
``When I'm driving myself, I probably wouldn't use ambient lighting at all---it's distracting. But as a passenger, I'd definitely want it.'' (P15)
\end{quote}

\begin{quote}
``In manual driving, I'd avoid any relaxing elements... But in autonomous mode, I'd welcome that relaxation.'' (P21)
\end{quote}

Eleven participants explicitly mentioned personalization needs:

\begin{quote}
``I think these should be completely customizable. What relaxes me might agitate someone else.'' (P24)
\end{quote}

These qualitative patterns validate the empirical findings while revealing individual-level mechanisms underlying the group-level statistical effects, providing convergent evidence for the atmospheric processing architecture identified through quantitative analysis.

\section{Discussion}
When passengers no longer operate vehicles but inhabit AI-controlled spaces, how do they experience and evaluate these intelligent systems? This study addressed this question by examining atmospheric design in autonomous vehicles. Three principal findings emerged. First, multisensory atmospheric interventions substantially influenced passenger experience,  with holistic atmospheric perception accounting for 66.5\% of variance in journey evaluation---an effect magnitude exceeding typical retail atmospheric effects. Second, atmospheric processing followed a hierarchical cascade from independent sensory appraisals through strategic cross-modal integration to unified environmental evaluation, with multisensory interactions emerging selectively at higher-order evaluative stages rather than basic perceptual levels. Third, olfactory stimuli consistently exerted stronger influence than visual stimuli across arousal, pleasure, and journey evaluation, reversing typical visual dominance in transport contexts. Collectively, these findings demonstrate that when operational control shifts from humans to AI, environmental quality becomes a primary rather than peripheral interface mediating passenger-system relationships. This has important implications for understanding human-AI interaction more broadly. We discuss mechanisms underlying environmental interfaces, specify atmospheric processing architecture in technology-mediated contexts, examine methodological considerations, and identify critical directions for future research.

\subsection{Theoretical Contributions and Implications}

\subsubsection{Environmental Interfaces in Human-AI Interaction}
This investigation extends atmospheric principles from stationary commercial environments to dynamic, confined autonomous vehicle contexts. Atmospheric perception emerged as the dominant pathway mediating passenger experience, accounting for 66.5\% of within-person variance in journey evaluation ($R^2_{Overall Experience} = .665$). This magnitude appears to exceed environmental effects typically reported in retail settings (15-30\%) \cite{Baker2002-lk, Morrison2011-uq}. While direct comparison requires caution given differences between within-person experimental designs and cross-sectional correlational studies, the strength of atmospheric effects suggests that unique characteristics of autonomous vehicle cabins may amplify rather than diminish environmental influence. Several factors may contribute to this amplification. First, spatial confinement: unlike retail environments where customers move freely, AV passengers inhabit a bounded space without alternative environmental options. Second, involuntary exposure: passengers cannot exit during travel, intensifying atmospheric impact. Third, the absence of operational control: when traditional interfaces are eliminated, environmental quality may assume primary importance as passengers seek experiential anchors for evaluating AI systems.

These findings challenge classic technology acceptance models where functional attributes drive evaluation \cite{Davis1989-ao}. When operational control transfers to AI systems, environmental quality shifts from supporting role to primary interface. We propose the concept of environmental interfaces: atmospheric conditions mediating human-technology relationships through ambient, holistic pathways rather than explicit, functional ones. This extends Dourish's embodied interaction framework \cite{Dourish2001-nq} to encompass environmental mediation of human-AI relationships. The theoretical implications extend beyond transportation. As AI systems assume operational control across domains (smart homes, automated workplaces, healthcare environments), design paradigms may need to shift from supporting direct manipulation \cite{Norman1988-pb} to cultivating environmental conditions supporting human inhabitation of intelligent systems. Understanding environmental interfaces as a distinct interaction modality represents an important theoretical contribution to human-AI interaction research.

\subsubsection{Hierarchical Processing Architecture for Atmospheric Experience}
Our pathway analysis provides mechanistic specificity for atmospheric processing in technology-mediated environments, revealing a clear progression from modality-specific sensory preferences through cross-modal integration to holistic atmospheric assessment. This hierarchical sequence offers empirical support for theoretical frameworks proposing staged environmental processing \cite{Russell1977-yu, gibson2014ecological} while providing the mechanistic detail previously lacking in autonomous vehicle contexts. A critical finding concerns the selective emergence of multisensory interactions. Lighting and scent effects operated independently at basic perceptual levels (sensory preferences showed no significant interactions) but exhibited significant interactions for higher-order outcomes (Trust, Overall Experience, Pleasure). This pattern suggests cross-modal integration occurs strategically at evaluative stages rather than automatically at perceptual stages. The finding challenges models proposing early sensory fusion \cite{Spence2011-pn} while supporting frameworks emphasizing context-dependent integration strategies \cite{Norman1975-bs}. Integration mechanisms appear dynamically deployed to meet higher-order evaluative demands rather than operating as obligatory perceptual processes.

Particularly significant is the finding that affective responses operate as parallel outcomes rather than sequential mediators of experience. While atmospheric appraisals systematically influenced emotional states (Perceived Atmosphere predicted Pleasure: $b=0.753$, $p<.001$; Arousal: $b=-0.246$, $p=.0019$), these emotional changes contributed negligibly to final experiential outcomes once cognitive atmospheric evaluation was considered. Adding affect measures to outcome models produced minimal incremental variance ($\Delta R^2_m \approx -0.002$), with joint Wald tests confirming negligible contribution (Trust model: $\chi^2(2)=0$, $p=1.00$). This pattern offers an important refinement of classic emotion-centered models of environmental influence \cite{Russell1977-yu} which position affect as the central mediator. In environments where the primary goal is hedonic (retail, hospitality), affect may indeed drive outcomes. However, in technology-mediated mobility contexts where passengers must evaluate opaque AI systems, cognitively-driven pathways predominate. The processing sequence appears to be: environmental stimuli influence cognitive-evaluative processes (passengers' active appraisal of atmospheric quality, comfort, coherence), which drive both evaluative judgments (Trust, Overall Experience) and parallel affective responses (Pleasure, Arousal). This refined S-O-R architecture represents an important theoretical contribution to applying environmental psychology in complex technology-mediated settings.

\subsubsection{Attentional Liberation and Sensory Hierarchy Reconfiguration}
A central finding is the systematic prominence of olfactory over visual effects, presenting a compelling counter-narrative to visual dominance typically assumed in automotive interface research \cite{Engstrom2017-yi}. Scent manipulations consistently produced larger effect sizes than lighting across arousal (scent $\eta^2_p=.320$ vs. lighting $\eta^2_p=.125$), pleasure (scent $\eta^2_p=.264$ vs. lighting $\eta^2_p=.131$), and overall experience (scent $\eta^2_p=.216$ vs. lighting $\eta^2_p=.127$). This pattern suggests fundamental reconfiguration of sensory hierarchies in autonomous vehicle contexts. We propose an attentional liberation effect to explain this reconfiguration. In driver-operated vehicles, visual attention must focus on safety-critical information (road conditions, traffic, instruments), constraining perceptual engagement with ambient environmental qualities. When AI assumes operational control, passengers are freed from these attentional demands, enabling fuller perceptual engagement with ambient modalities\cite{eddine2024investigating}. Olfactory stimuli may particularly benefit from this liberation because scent operates through direct limbic pathways that do not require focal attentional allocation \cite{Zald1997-wm, Herz2009-rn}. In active-engagement contexts (retail, dining), focal visual elements naturally dominate because they align with task demands. In passive-inhabitation contexts (autonomous vehicles), ambient modalities like scent can exert disproportionate influence through peripheral processing channels. This interpretation implies that sensory hierarchies are task-contingent rather than fixed, reconfiguring based on operational demands and engagement modes. The finding challenges design principles forged under human operational constraints. Visual interface design prioritizing focal attention may prove less effective when users inhabit rather than operate intelligent systems. As AI systems assume operational control across domains, design paradigms developed for human operation may require fundamental reconsideration. The attentional liberation effect represents an important theoretical contribution with broad implications for human-AI interface design beyond autonomous vehicles.

\subsubsection{Environmental Pathways to Technology Trust}
Our findings address the evaluation gap inherent in opaque AI systems by identifying two distinct environmental pathways to technology trust. First, a primary route through holistic atmospheric perception ($\beta=0.494$, $p<.001$), accounting for 67.1\% of variance in Trust. Second, a secondary route through sensory congruency evaluation ($\beta=0.169$, $p=.037$), contributing additional 2.8\% variance. These pathways operate independently of traditional performance-based trust mechanisms \cite{Lee2004-uz}. The atmospheric perception pathway suggests passengers use environmental quality as a proxy signal for system sophistication when direct performance assessment proves difficult. High-quality, coherent atmospheres function as cues for system reliability and competence \cite{Spence2020-bu}. This mechanism extends signaling theory \cite{Spence1973-aa} to human-AI contexts while providing empirical support for the aesthetic-usability effect \cite{Tractinsky2000-ej} in safety-critical domains where trust assumes heightened importance. The sensory congruency pathway operates through coherence evaluation. Qualitative data revealed passengers interpret environmental mismatches as indicating system-level incoherence: "If the colors and scents feel mismatched, it makes me question whether the system is coherent... it affects my confidence in the technology" (P18). This symbolic evaluation process suggests passengers extend Gestalt principles from environmental perception to technology assessment, inferring system competence from environmental design quality. Critically, trust operated as a parallel outcome rather than mediating overall experience. Nested model comparisons showed removing the trust$\rightarrow$experience path did not degrade model fit (Satorra-Bentler $\Delta\chi^2(2)=2.79$, $p=.248$), indicating atmospheric quality influences trust and experience through separate psychological processes. This pattern suggests environmental trust formation represents a distinct evaluation dimension rather than an intermediate step toward experiential outcomes. These findings make important theoretical contributions to trust research in autonomous systems. They demonstrate that trust can form through environmental pathways when traditional performance-based mechanisms are unavailable, and they identify specific mechanisms (atmospheric quality, sensory congruency) through which environmental design influences trust formation. Understanding environmental trust pathways proves particularly important for AI systems where algorithmic opacity prevents direct competence assessment.

\subsubsection{Compensatory Multisensory Integration Dynamics}
The discovery of compensatory rather than additive multisensory effects provides new insights into atmospheric design optimization. Olfactory benefits were strongest under suboptimal lighting conditions: for Trust, mint and lavender enhanced ratings under orange lighting ($p=.002$ and $p=.011$) but showed no differential effects under blue lighting; for Overall Experience, scent benefits emerged under orange and no lighting but were eliminated under blue lighting. This pattern suggests atmospheric elements function as substitutable resources where adequacy in one modality reduces influence of additional inputs. This compensatory mechanism challenges traditional atmospheric design approaches assuming proportional enhancement across sensory channels \cite{Kotler1973-es, Bitner1992-xe}. Instead, findings suggest adequacy- based resource allocation where environmental satisfaction operates through threshold rather than maximization principles. Once satisfactory atmospheric baseline is established through one modality, additional sensory inputs contribute marginally to experiential outcomes. The compensatory pattern aligns with cognitive resource allocation theories \cite{Norman1975-bs, Wickens2020-xm} demonstrating dynamic processing strategy adjustment based on situational demands. The theoretical contribution lies in demonstrating that multisensory integration in atmospheric contexts follows strategic resource allocation principles rather than simple additive combination. Practical implications suggest efficient atmospheric design may focus on achieving threshold quality in one or two modalities rather than maximizing all sensory dimensions simultaneously, particularly relevant given energy and sustainability constraints in vehicle environments.

\subsection{Methodological Contributions and Limitations}
\subsubsection{Experimental Design Strengths}
The within-subjects factorial design provided rigorous causal inference while maximizing statistical efficiency. The resulting 216 observations (24 participants $\times$ 9 conditions) generated substantial power for detecting atmospheric effects, with observed effect sizes ($\eta^2_p$ range: .114-.320) confirming robust manipulation effectiveness. Latin square counterbalancing controlled order effects while enabling systematic exposure to all atmospheric combinations, particularly valuable given individual differences in atmospheric sensitivity. 

The integration of experimental manipulation with pathway modeling represents a methodological contribution to atmospheric research. While most studies employ either experimental or correlational approaches, our sequential strategy (ANOVA followed by pathway modeling) enabled both causal inference and mechanistic specification. This approach proved particularly effective for exploratory research in emerging technological contexts where established theoretical relationships require empirical validation.

\subsubsection{Limitations and Constraints}
However, the intensive within-subjects approach raises ecological validity questions. While rest periods and cognitive distraction tasks followed established protocols for minimizing carryover effects \cite{Otto2021-zm}, the compressed exposure sequence differs substantially from naturalistic atmospheric experience patterns. The 75-second exposure periods represent methodological compromise between capturing initial atmospheric responses and avoiding adaptation effects, sufficient for detecting systematic effects \cite{Hummel2007-mb} but inadequate for addressing temporal dynamics characterizing extended journeys. 

Laboratory-based VR simulation enabled precise atmospheric control while providing immersive experiences. The Tesla Model 3 interior replication with controlled temperature and humidity maintained consistent conditions for olfactory perception \cite{Sela2010-qy}. Nevertheless, simulation involves inherent tradeoffs between experimental control and ecological realism. Real-world autonomous vehicle experiences involve dynamic environmental conditions, social contexts, journey purposes, and extended temporal patterns that controlled laboratory conditions cannot fully represent. Our approach provided proof-of-concept evidence and mechanism identification while establishing need for naturalistic validation. 

The sample (predominantly younger adults, $M=29.6$ years, with driving experience) aligns with early autonomous vehicle adopter demographics while potentially constraining generalizability. The within-subjects design partially mitigates representativeness concerns by focusing on individual response patterns rather than population estimates, though cross-cultural validation remains important given documented variation in sensory preferences and environmental values \cite{Hofstede1988-fb}. 

The exploratory nature of this investigation makes traditional power calculations somewhat inappropriate, as we sought to identify atmospheric mechanisms rather than test predetermined hypotheses. While larger samples would enable more sophisticated modeling approaches (confirmatory factor analysis, multilevel structural equation models), the current sample provided adequate power for detecting medium-to-large effects characterizing environmental interventions while enabling intensive within-subjects manipulations necessary for mechanism identification.

\subsection{Implications for Design and Practice}
Findings suggest several design principles for autonomous vehicle atmospheres with broader implications for human-AI interaction. First, holistic environmental quality should be considered a primary interface rather than secondary aesthetic consideration. Atmospheric perception accounts for substantial variance in both journey satisfaction (66.5\%) and technology trust (67.1\%), warranting systematic attention in vehicle design processes and resource allocation decisions.

Second, olfactory design deserves greater consideration than typically allocated in automotive contexts. The consistent olfactory prominence across arousal, pleasure, and overall experience suggests ambient scent may prove more influential than visual ambient lighting for passenger experience. This challenges current design priorities emphasizing visual elements while suggesting strategic reallocation toward olfactory interventions, particularly given the direct limbic pathways through which scent operates.

Third, compensatory integration dynamics suggest efficient atmospheric design may focus on achieving threshold quality in one or two modalities rather than maximizing all sensory dimensions simultaneously. This principle proves particularly relevant given energy and sustainability constraints in vehicle environments. Design strategies might prioritize establishing satisfactory baseline conditions through cost-effective modalities while reserving intensive interventions for contexts where primary channels prove inadequate.

Fourth, environmental quality influences technology trust through distinct pathways from traditional performance-based mechanisms. This finding suggests atmospheric design represents not merely experiential enhancement but trust-building strategy for AI systems where direct competence assessment proves difficult. Design practices should recognize environmental quality as trust signal, with implications extending beyond transportation to any context where AI assumes operational control.

\subsection{Future Research Directions}
\subsubsection{Naturalistic Validation and Extended Exposure Studies}
The transition from laboratory proof-of-concept to naturalistic validation represents the most critical next step. Field studies in actual autonomous vehicles over extended journey periods will determine whether atmospheric processing mechanisms identified here operate consistently under real-world conditions characterized by environmental variability, social complexity, and temporal dynamics. 

Longitudinal investigations are particularly needed to understand adaptation patterns and preference evolution. Our brief exposures avoided adaptation effects, but extended atmospheric exposure may involve habituation, preference changes, or temporal patterns not captured in laboratory conditions. Research on environmental adaptation suggests complex dynamics where some atmospheric effects diminish with repeated exposure while others strengthen through association and familiarity \cite{Bornstein1975-jk}. Understanding these temporal patterns proves essential for designing effective long-term atmospheric interventions.

\subsubsection{Goal-Congruent Atmospheric Design}
While this investigation established general atmospheric influence mechanisms, autonomous vehicle cabins serve as platforms for diverse passenger activities (working, resting, socializing)~\cite{-_Fagnant2015-af, -_Pettigrew2023-he}. Future research should investigate goal-congruent atmospheric design: how environmental stimuli can be orchestrated to shape specific atmospheric profiles (quiet focus for work, serene tranquility for rest) optimizing activity-specific outcomes (productivity, relaxation quality, sleep quality). Such research would represent significant progression from foundational principles to applied, personalized passenger-centric design.

\subsubsection{Individual Differences and Adaptive Systems}
Understanding individual differences in atmospheric processing represents both theoretical and practical importance. Factors including sensory processing sensitivity \cite{Aron1997-cc}, cultural background \cite{Krishna2012-qq}, and personal associations \cite{Herz2004-yt} likely moderate atmospheric mechanisms. Qualitative data revealed substantial individual variation in sensory prioritization strategies, with some participants emphasizing olfactory inescapability while others prioritized visual continuity. Investigation of adaptive atmospheric systems responding to individual preferences while maintaining effective environmental design principles could enable personalized environmental optimization.

\section{Conclusion}
As AI systems assume operational control across domains, understanding how humans experience and evaluate intelligent environments becomes increasingly critical. This study examined atmospheric design in autonomous vehicles---a context where traditional control interfaces are eliminated and passengers must evaluate opaque AI systems through environmental cues. Results demonstrate that multisensory atmospheric interventions substantially influence both journey satisfaction and technology trust, operating through hierarchical processing mechanisms from sensory appraisals to holistic environmental evaluation.

Four theoretical contributions emerge. First, we establish environmental quality as a primary rather than peripheral interface when operational control transfers to AI, with atmospheric perception accounting for 66.5\% of variance in journey evaluation. This magnitude exceeds typical retail atmospheric effects, suggesting amplification in confined, technology-mediated contexts. Second, we specify hierarchical atmospheric processing where multisensory integration occurs strategically at evaluative stages rather than automatically at perceptual stages, refining classic S-O-R frameworks for technology-mediated environments. Third, we identify systematic olfactory dominance over visual effects---a reversal of typical transport sensory hierarchies attributable to attentional liberation when operational demands are removed. Fourth, we demonstrate that environmental quality influences technology trust through distinct pathways (atmospheric perception, sensory congruency) operating independently of traditional performance-based mechanisms.

These findings have implications beyond autonomous vehicles. As AI assumes operational roles in smart homes, automated workplaces, and intelligent healthcare environments, design paradigms may need to shift from supporting direct manipulation to cultivating environmental conditions that support human inhabitation of intelligent systems. Understanding environmental interfaces as a fundamental rather than peripheral aspect of human-AI interaction represents an important theoretical contribution with practical implications for designing symbiotic human-AI environments. Future research should validate these mechanisms in naturalistic settings, investigate goal-congruent atmospheric design for diverse passenger activities, and examine how individual differences and cultural backgrounds moderate atmospheric processing in technology-mediated contexts.

\appendix

\section*{Supplementary Material}

\renewcommand{\thesubsection}{S\arabic{subsection}}

\subsection{rm-ANOVA Post-hoc Comparisons (R/afex + emmeans)}\label{supp:S1}

Post-hoc tests are reported only when the corresponding omnibus effect is significant ($\alpha = .05$, Greenhouse--Geisser corrected where applicable). If the interaction (Light $\times$ Scent) is significant, simple effects are reported (Light within each level of Scent and/or Scent within each level of Light). If only a main effect is significant, pairwise comparisons within that factor (collapsed across the other factor) are reported. No post-hoc tests are provided when the omnibus effect is non-significant. All $p$-values in this section are Bonferroni-adjusted within each family of comparisons.

\subsubsection{Sensory preference\_light (person-mean centered)}
Warranted post-hoc: Light main effect (significant); Scent (ns); Interaction (ns). Reported: pairwise Light contrasts only.

\begin{table}[H]
\centering
\caption{Sensory preference\_light --- pairwise Light contrasts (Bonferroni-adjusted).}
\label{tab:S1-1}
\begin{threeparttable}
\begin{tabular}{lccccc}
\toprule
Contrast & Estimate & SE & df & $t$ & $p_{\text{adj}}$ \\
\midrule
None -- Blue   & $-1.139$ & 0.399 & 23 & $-2.856$ & .0268 \\
None -- Orange & 0.111    & 0.333 & 23 & 0.334    & 1 \\
Blue -- Orange & 1.250    & 0.447 & 23 & 2.794    & .0309 \\
\bottomrule
\end{tabular}
\begin{tablenotes}
\item Note. Scent post-hocs omitted (omnibus ns).
\end{tablenotes}
\end{threeparttable}
\end{table}

\subsubsection{Sensory preference\_scent (person-mean centered)}
Warranted post-hoc: Scent main effect (significant); Light (ns); Interaction (ns). Reported: pairwise Scent contrasts only.

\begin{table}[H]
\centering
\caption{Sensory preference\_scent --- pairwise Scent contrasts (Bonferroni-adjusted).}
\label{tab:S1-2}
\begin{threeparttable}
\begin{tabular}{lccccc}
\toprule
Contrast & Estimate & SE & df & $t$ & $p_{\text{adj}}$ \\
\midrule
None -- Mint     & $-0.8750$ & 0.263 & 23 & $-3.327$ & .0088 \\
None -- Lavender & 0.0556    & 0.445 & 23 & 0.125    & 1 \\
Mint -- Lavender & 0.9306    & 0.416 & 23 & 2.239    & .1053 \\
\bottomrule
\end{tabular}
\begin{tablenotes}
\item Note. Light post-hocs omitted (omnibus ns).
\end{tablenotes}
\end{threeparttable}
\end{table}

\subsubsection{Sensory preference\_combination (person-mean centered)}
Warranted post-hoc: Light main effect (significant); Scent (marginal, ns); Interaction (ns). Reported: pairwise Light contrasts only.

\begin{table}[H]
\centering
\caption{Sensory preference\_combination --- pairwise Light contrasts (Bonferroni-adjusted).}
\label{tab:S1-3}
\begin{tabular}{lccccc}
\toprule
Contrast & Estimate & SE & df & $t$ & $p_{\text{adj}}$ \\
\midrule
None -- Blue   & $-0.472$ & 0.275 & 23 & $-1.717$ & .2983 \\
None -- Orange & 0.319    & 0.225 & 23 & 1.421    & .5059 \\
Blue -- Orange & 0.792    & 0.307 & 23 & 2.575    & .0508 \\
\bottomrule
\end{tabular}
\end{table}

\subsubsection{Perceived Atmosphere (person-mean centered)}
Warranted post-hoc: Light (significant), Scent (significant), Interaction (marginal, ns). Reported: pairwise contrasts for Light and Scent.

\begin{table}[H]
\centering
\caption{Perceived Atmosphere --- pairwise Light contrasts (Bonferroni-adjusted).}
\label{tab:S1-4A}
\begin{threeparttable}
\begin{tabular}{lccccc}
\toprule
Contrast & Estimate & SE & df & $t$ & $p_{\text{adj}}$ \\
\midrule
None -- Blue   & $-0.532$ & 0.191 & 23 & $-2.785$ & .0316 \\
None -- Orange & 0.204    & 0.227 & 23 & 0.899    & 1 \\
Blue -- Orange & 0.736    & 0.271 & 23 & 2.716    & .0369 \\
\bottomrule
\end{tabular}
\end{threeparttable}
\end{table}

\begin{table}[H]
\centering
\caption{Perceived Atmosphere --- pairwise Scent contrasts (Bonferroni-adjusted).}
\label{tab:S1-4B}
\begin{threeparttable}
\begin{tabular}{lccccc}
\toprule
Contrast & Estimate & SE & df & $t$ & $p_{\text{adj}}$ \\
\midrule
None -- Mint     & $-0.792$ & 0.208 & 23 & $-3.798$ & .0028 \\
None -- Lavender & $-0.120$ & 0.280 & 23 & $-0.429$ & 1 \\
Mint -- Lavender & 0.671    & 0.266 & 23 & 2.521    & .0573 \\
\bottomrule
\end{tabular}
\end{threeparttable}
\end{table}

\subsubsection{Sensory Congruency (person-mean centered)}
Warranted post-hoc: Light (significant), Scent (significant), Interaction (ns). Reported: pairwise contrasts for Light and Scent.

\begin{table}[H]
\centering
\caption{Sensory Congruency --- pairwise Light contrasts (Bonferroni-adjusted).}
\label{tab:S1-5A}
\begin{tabular}{lccccc}
\toprule
Contrast & Estimate & SE & df & $t$ & $p_{\text{adj}}$ \\
\midrule
None -- Blue   & 0.167  & 0.215 & 23 & 0.774 & 1 \\
None -- Orange & 0.688  & 0.180 & 23 & 3.829 & .0026 \\
Blue -- Orange & 0.521  & 0.230 & 23 & 2.263 & .1002 \\
\bottomrule
\end{tabular}
\end{table}

\begin{table}[H]
\centering
\caption{Sensory Congruency --- pairwise Scent contrasts (Bonferroni-adjusted).}
\label{tab:S1-5B}
\begin{tabular}{lccccc}
\toprule
Contrast & Estimate & SE & df & $t$ & $p_{\text{adj}}$ \\
\midrule
None -- Mint     & 0.0139 & 0.145 & 23 & 0.096 & 1 \\
None -- Lavender & 0.5069 & 0.206 & 23 & 2.461 & .0653 \\
Mint -- Lavender & 0.4931 & 0.195 & 23 & 2.530 & .0562 \\
\bottomrule
\end{tabular}
\end{table}

\subsubsection{Pleasure Change (person-mean centered)}
Warranted post-hoc: Interaction (significant). Reported: simple effects both ways.

\begin{table}[H]
\centering
\caption{Pleasure --- simple effects of Light within each Scent (Bonferroni-adjusted).}
\label{tab:S1-6A}
\begin{tabular}{llccccc}
\toprule
Scent & Contrast & Estimate & SE & df & $t$ & $p_{\text{adj}}$ \\
\midrule
None     & none -- blue   & $-0.8750$ & 0.297 & 23 & $-2.948$ & .0216 \\
         & none -- orange & 0.2917    & 0.252 & 23 & 1.159    & .7751 \\
         & blue -- orange & 1.1667    & 0.339 & 23 & 3.444    & .0066 \\
Mint     & none -- blue   & 0.2083    & 0.307 & 23 & 0.679    & 1 \\
         & none -- orange & 0.0833    & 0.275 & 23 & 0.303    & 1 \\
         & blue -- orange & $-0.1250$ & 0.326 & 23 & $-0.384$ & 1 \\
Lavender & none -- blue   & $-0.5417$ & 0.318 & 23 & $-1.701$ & .3071 \\
         & none -- orange & 0.25      & 0.308 & 23 & 0.811    & 1 \\
         & blue -- orange & 0.7917    & 0.493 & 23 & 1.607    & .3651 \\
\bottomrule
\end{tabular}
\end{table}

\begin{table}[H]
\centering
\caption{Pleasure --- simple effects of Scent within each Light (Bonferroni-adjusted).}
\label{tab:S1-6B}
\begin{tabular}{llccccc}
\toprule
Light & Contrast & Estimate & SE & df & $t$ & $p_{\text{adj}}$ \\
\midrule
None   & none -- mint     & $-1.2083$ & 0.248 & 23 & $-4.872$ & .0002 \\
       & none -- lavender & $-0.0833$ & 0.408 & 23 & $-0.204$ & 1 \\
       & mint -- lavender & 1.125     & 0.401 & 23 & 2.808    & .0300 \\
Blue   & none -- mint     & $-0.1250$ & 0.368 & 23 & $-0.340$ & 1 \\
       & none -- lavender & 0.25      & 0.396 & 23 & 0.632    & 1 \\
       & mint -- lavender & 0.375     & 0.365 & 23 & 1.027    & .9456 \\
Orange & none -- mint     & $-1.4167$ & 0.306 & 23 & $-4.623$ & .0004 \\
       & none -- lavender & $-0.1250$ & 0.315 & 23 & $-0.397$ & 1 \\
       & mint -- lavender & 1.2917    & 0.310 & 23 & 4.170    & .0011 \\
\bottomrule
\end{tabular}
\end{table}

\subsubsection{Arousal Change (person-mean centered)}
Warranted post-hoc: Scent main effect (significant). Reported: pairwise Scent contrasts only.

\begin{table}[H]
\centering
\caption{Arousal --- pairwise Scent contrasts (Bonferroni-adjusted).}
\label{tab:S1-7}
\begin{tabular}{lccccc}
\toprule
Contrast & Estimate & SE & df & $t$ & $p_{\text{adj}}$ \\
\midrule
None -- Mint     & $-1.028$ & 0.244 & 23 & $-4.212$ & .001 \\
None -- Lavender & $-0.847$ & 0.255 & 23 & $-3.328$ & .0088 \\
Mint -- Lavender & 0.181    & 0.206 & 23 & 0.878    & 1 \\
\bottomrule
\end{tabular}
\end{table}

\subsubsection{Overall Experience (person-mean centered)}
Warranted post-hoc: Interaction (significant). Reported: simple effects both ways.

\begin{table}[H]
\centering
\caption{Overall Experience --- simple effects of Light within each Scent (Bonferroni-adjusted).}
\label{tab:S1-8A}
\begin{tabular}{llccccc}
\toprule
Scent & Contrast & Estimate & SE & df & $t$ & $p_{\text{adj}}$ \\
\midrule
None     & none -- blue   & $-1.3333$ & 0.398 & 23 & $-3.352$ & .0083 \\
         & none -- orange & 0.0833    & 0.294 & 23 & 0.283    & 1 \\
         & blue -- orange & 1.4167    & 0.361 & 23 & 3.927    & .0020 \\
Mint     & none -- blue   & 0         & 0.430 & 23 & 0        & 1 \\
         & none -- orange & 0.375     & 0.317 & 23 & 1.181    & .7486 \\
         & blue -- orange & 0.375     & 0.458 & 23 & 0.819    & 1 \\
Lavender & none -- blue   & $-0.5417$ & 0.335 & 23 & $-1.617$ & .3587 \\
         & none -- orange & $-0.4583$ & 0.351 & 23 & $-1.306$ & .6131 \\
         & blue -- orange & 0.0833    & 0.541 & 23 & 0.154    & 1 \\
\bottomrule
\end{tabular}
\end{table}

\begin{table}[H]
\centering
\caption{Overall Experience --- simple effects of Scent within each Light (Bonferroni-adjusted).}
\label{tab:S1-8B}
\begin{tabular}{llccccc}
\toprule
Light & Contrast & Estimate & SE & df & $t$ & $p_{\text{adj}}$ \\
\midrule
None   & none -- mint     & $-1.4583$ & 0.340 & 23 & $-4.284$ & .0008 \\
       & none -- lavender & $-0.4583$ & 0.462 & 23 & $-0.991$ & .9953 \\
       & mint -- lavender & 1.0000    & 0.319 & 23 & 3.140    & .0138 \\
Blue   & none -- mint     & $-0.1250$ & 0.315 & 23 & $-0.397$ & 1 \\
       & none -- lavender & 0.3333    & 0.354 & 23 & 0.941    & 1 \\
       & mint -- lavender & 0.4583    & 0.514 & 23 & 0.891    & 1 \\
Orange & none -- mint     & $-1.1667$ & 0.374 & 23 & $-3.117$ & .0145 \\
       & none -- lavender & $-1.0000$ & 0.385 & 23 & $-2.595$ & .0486 \\
       & mint -- lavender & 0.1667    & 0.305 & 23 & 0.547    & 1 \\
\bottomrule
\end{tabular}
\end{table}

\subsubsection{Trust (person-mean centered)}
Warranted post-hoc: Interaction (significant). Reported: simple effects both ways.

\begin{table}[H]
\centering
\caption{Trust --- simple effects of Light within each Scent (Bonferroni-adjusted).}
\label{tab:S1-9A}
\begin{tabular}{llccccc}
\toprule
Scent & Contrast & Estimate & SE & df & $t$ & $p_{\text{adj}}$ \\
\midrule
None     & none -- blue   & $-0.6250$ & 0.317 & 23 & $-1.969$ & .1834 \\
         & none -- orange & 0.6667    & 0.291 & 23 & 2.289    & .0948 \\
         & blue -- orange & 1.2917    & 0.360 & 23 & 3.590    & .0046 \\
Mint     & none -- blue   & $-0.2292$ & 0.261 & 23 & $-0.879$ & 1 \\
         & none -- orange & 0.0417    & 0.246 & 23 & 0.169    & 1 \\
         & blue -- orange & 0.2708    & 0.262 & 23 & 1.032    & .9379 \\
Lavender & none -- blue   & $-0.1250$ & 0.280 & 23 & $-0.447$ & 1 \\
         & none -- orange & 0.1875    & 0.269 & 23 & 0.697    & 1 \\
         & blue -- orange & 0.3125    & 0.377 & 23 & 0.829    & 1 \\
\bottomrule
\end{tabular}
\end{table}

\begin{table}[H]
\centering
\caption{Trust --- simple effects of Scent within each Light (Bonferroni-adjusted).}
\label{tab:S1-9B}
\begin{tabular}{llccccc}
\toprule
Light & Contrast & Estimate & SE & df & $t$ & $p_{\text{adj}}$ \\
\midrule
None   & none -- mint     & $-0.4375$ & 0.275 & 23 & $-1.593$ & .3747 \\
       & none -- lavender & $-0.2917$ & 0.287 & 23 & $-1.016$ & .9601 \\
       & mint -- lavender & 0.1458    & 0.280 & 23 & 0.521    & 1 \\
Blue   & none -- mint     & $-0.0417$ & 0.173 & 23 & $-0.241$ & 1 \\
       & none -- lavender & 0.2083    & 0.244 & 23 & 0.853    & 1 \\
       & mint -- lavender & 0.2500    & 0.227 & 23 & 1.100    & .8478 \\
Orange & none -- mint     & $-1.0625$ & 0.294 & 23 & $-3.616$ & .0044 \\
       & none -- lavender & $-0.7708$ & 0.239 & 23 & $-3.227$ & .0112 \\
       & mint -- lavender & 0.2917    & 0.317 & 23 & 0.920    & 1 \\
\bottomrule
\end{tabular}
\end{table}

\subsection{Exploratory within-only SEM (triangulation) --- Tables}\label{supp:S2}

\noindent\textit{Modeling note.} Person-mean centered endogenous variables; treatment coding (reference = ``None''); latent \emph{Perceived Atmosphere}, \emph{Sensory Congruency}, \emph{Trust}; manifest \emph{Sensory preference\_light}, \emph{Sensory preference\_scent}, \emph{Sensory preference\_combination}, \emph{Overall Experience}. Estimator: MLR with FIML; participant-clustered SEs. With $N=24$, SEM is not used for confirmatory inference; it triangulates the serial mixed-model results and provides a single-figure overview.

\begin{table}[H]
\centering
\caption{Key standardized paths (Std.all) and $R^2$ (SEM; within-only).}
\label{tab:SOR-1}
\begin{threeparttable}
\begin{tabular}{lcc}
\toprule
From $\rightarrow$ To & Std.all & $p$ \\
\midrule
\multicolumn{3}{l}{\textbf{Stimuli $\rightarrow$ Single-channel preference (O$_1$)}}\\
Blue $\rightarrow$ Sensory preference\_light & 0.378 & .003 \\
Orange $\rightarrow$ Sensory preference\_light & $-0.033$ & .761 \\
Mint $\rightarrow$ Sensory preference\_scent & 0.292 & .001 \\
Lavender $\rightarrow$ Sensory preference\_scent & $-0.025$ & .867 \\
\addlinespace[2pt]
\multicolumn{3}{l}{\textbf{Single-channel $\rightarrow$ Combination}}\\
Sensory preference\_light $\rightarrow$ Sensory preference\_combination & 0.594 & $<.001$ \\
Sensory preference\_scent $\rightarrow$ Sensory preference\_combination & 0.476 & $<.001$ \\
\addlinespace[2pt]
\multicolumn{3}{l}{\textbf{Combination/Channels $\rightarrow$ Appraisals (O$_2$/O$_3$)}}\\
Sensory preference\_combination $\rightarrow$ Sensory Congruency & 0.732 & $<.001$ \\
Sensory preference\_combination $\rightarrow$ Perceived Atmosphere & 0.486 & $<.001$ \\
Sensory preference\_light $\rightarrow$ Perceived Atmosphere & 0.258 & $<.001$ \\
Sensory preference\_scent $\rightarrow$ Perceived Atmosphere & 0.407 & $<.001$ \\
\addlinespace[2pt]
\multicolumn{3}{l}{\textbf{Appraisals/Stimuli $\rightarrow$ Emotions}}\\
Perceived Atmosphere $\rightarrow$ Pleasure & 0.767 & $<.001$ \\
Perceived Atmosphere $\rightarrow$ Arousal & $-0.257$ & .041 \\
Mint $\rightarrow$ Pleasure & 0.206 & $<.001$ \\
Mint $\rightarrow$ Arousal & 0.255 & $<.001$ \\
Blue$\times$Mint $\rightarrow$ Pleasure & $-0.123$ & .032 \\
\addlinespace[2pt]
\multicolumn{3}{l}{\textbf{Appraisals $\rightarrow$ Trust / Outcomes (R)}}\\
Perceived Atmosphere $\rightarrow$ Trust & 0.494 & $<.001$ \\
Sensory Congruency $\rightarrow$ Trust & 0.169 & .037 \\
Perceived Atmosphere $\rightarrow$ Overall Experience & 0.744 & $<.001$ \\
Trust $\rightarrow$ Overall Experience & 0.110$^\dagger$ & .079 \\
\bottomrule
\end{tabular}
\begin{tablenotes}
\item $^\dagger$ Retained for completeness; nested Satorra--Bentler $\Delta\chi^2(2)=2.79$, $p=.248$.
\item $R^2$: Sensory preference\_light = .156; Sensory preference\_scent = .094; Sensory preference\_combination = .684; Sensory Congruency = .536; Perceived Atmosphere = .910; Pleasure = .643; Arousal = .106; Trust = .383; Overall Experience = .665.
\end{tablenotes}
\end{threeparttable}
\end{table}

\begin{table}[H]
\centering
\caption{Total indirect effects of stimuli on Overall Experience (cluster bootstrap, 5{,}000).}
\label{tab:SOR-2}
\begin{tabular}{lccccc}
\toprule
Path & Estimate & SE & 95\% CI (lower) & 95\% CI (upper) & Std.all \\
\midrule
Blue $\rightarrow$ Overall Experience (total indirect) & 0.354 & 0.076 & 0.212 & 0.509 & 0.168 \\
Mint $\rightarrow$ Overall Experience (total indirect) & 0.318 & 0.081 & 0.164 & 0.482 & 0.151 \\
\bottomrule
\end{tabular}
\end{table}

\begin{table}[H]
\centering
\caption{Decomposition (via-Trust) and total effects to Trust (cluster bootstrap, 5{,}000).}
\label{tab:SOR-3}
\begin{threeparttable}
\begin{tabular}{lcccc}
\toprule
Quantity & Estimate & SE & 95\% CI (lower) & 95\% CI (upper) \\
\midrule
via Trust only (Blue $\rightarrow$ Overall Experience) & 0.030 & 0.017 & 0.001 & 0.070 \\
via Trust only (Mint $\rightarrow$ Overall Experience) & 0.025 & 0.015 & 0.001 & 0.060 \\
Blue $\rightarrow$ Trust (total) & 0.283 & 0.069 & 0.163 & 0.435 \\
Mint $\rightarrow$ Trust (total) & 0.239 & 0.068 & 0.120 & 0.383 \\
\bottomrule
\end{tabular}
\begin{tablenotes}
\item \textit{Global fit (context):} CFI = .924, TLI = .899, RMSEA = .093, SRMR = .049; robust CFI = .935, robust TLI = .913, robust RMSEA = .085.
\end{tablenotes}
\end{threeparttable}
\end{table}

\begin{figure}[H]
\centering
\includegraphics[width=.85\linewidth]{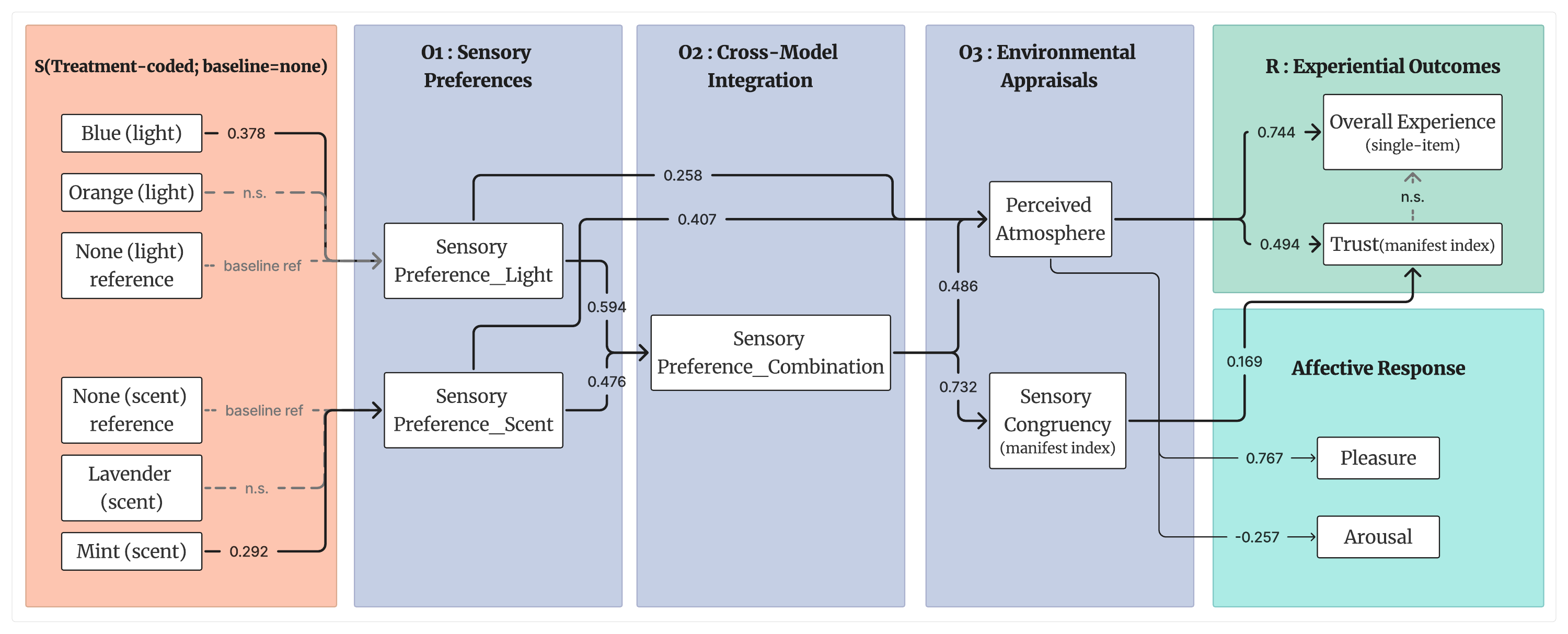}
\caption{Structural-only S--O--R model (within-only SEM; Std.all).}
\caption*{\textit{Note.} Manifest nodes: \emph{Sensory preference\_light}, \emph{Sensory preference\_scent}, \emph{Sensory preference\_combination}, \emph{Pleasure}, \emph{Arousal}, \emph{Overall Experience}. Latent nodes (ellipses): \emph{Perceived Atmosphere}, \emph{Sensory Congruency}, \emph{Trust}. Directed edges are fully standardized coefficients (line width $\propto |\beta|$). Solid = $p < .05$; dashed = retained but non-significant (e.g., Trust $\rightarrow$ Overall Experience). Stimulus $\rightarrow$ preference paths are deviations from the ``None'' baseline; Blue $\rightarrow$ Sensory preference\_light and Mint $\rightarrow$ Sensory preference\_scent are significant, Orange and Lavender are displayed as dashed for completeness. The Blue $\times$ Mint control is included in the Pleasure equation. Direct links from Pleasure/Arousal to Overall Experience were examined in a supplemental variant and were not retained for parsimony (see Supplement Table SOR-E1).}
\label{fig:SEM-structural}
\end{figure}

\begin{figure}[H]
\centering
\includegraphics[width=.85\linewidth]{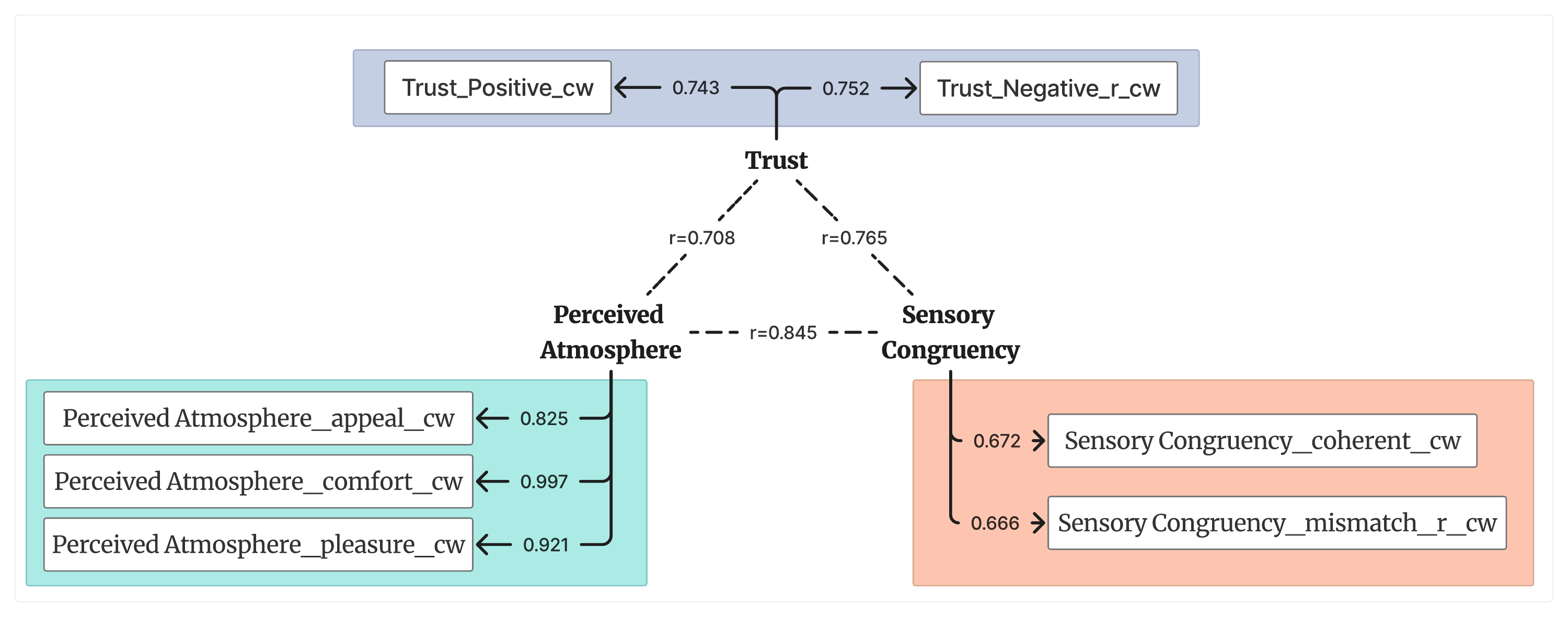}
\caption{Full measurement-structural diagram (Std.all).}
\caption*{\textit{Note.} As in Fig.~\ref{fig:SEM-structural}, with indicator loadings shown for \emph{Perceived Atmosphere} (appeal, comfort, pleasure), \emph{Sensory Congruency} (coordination; reverse-keyed intrusiveness), and \emph{Trust} (positive; reverse-keyed negative). Single-item nodes are drawn as rectangles without indicators by design. Indicator residuals omitted for readability.}
\label{fig:SEM-full}
\end{figure}

\subsection{Indirect effects via participant-cluster bootstrap (fixed-effects OLS)}\label{supp:S4}

For comparison with SEM (Tables~\ref{tab:SOR-2}--\ref{tab:SOR-3}), we also estimated total indirect effects using participant fixed-effects OLS with participant-cluster bootstrap ($B=5{,}000$). Results are directionally consistent.

\begin{table}[H]
\centering
\caption{Fixed-effects OLS with participant-cluster bootstrap (5{,}000).}
\label{tab:S4}
\begin{tabular}{lccc}
\toprule
Path/Quantity & Estimate & SE & $p$ (cluster) \\
\midrule
Blue $\rightarrow$ Overall Experience (total indirect) & 0.445 & 0.154 & .005 \\
Mint $\rightarrow$ Overall Experience (total indirect) & 0.439 & 0.150 & $<.001$ \\
via Trust only (Blue $\rightarrow$ Overall Experience) & 0.059 & 0.038 & .023 \\
via Trust only (Mint $\rightarrow$ Overall Experience) & 0.053 & 0.031 & .019 \\
Blue $\rightarrow$ Trust (total) & 0.259 & 0.100 & .005 \\
Mint $\rightarrow$ Trust (total) & 0.231 & 0.085 & $<.001$ \\
\bottomrule
\end{tabular}
\end{table}

\subsection{Affect as parallel responses}\label{supp:S5}

\subsubsection{Appraisal-to-Affect relationships (within-person regressions)}
\begin{table}[H]
\centering
\caption{Within-person regressions from appraisals to affect.}
\label{tab:S5-1}
\begin{tabular}{lccccc}
\toprule
Outcome & $b$ & SE & df & $t$ & $p$ \\
\midrule
Pleasure (cw) & 0.753 & 0.0481 & 214 & 15.7 & $2.36\times 10^{-37}$ \\
Arousal (cw)  & $-0.246$ & 0.0781 & 214 & $-3.15$ & .00187 \\
\bottomrule
\end{tabular}
\end{table}

\subsubsection{Incremental value of affect for Overall Experience and Trust}
\noindent Overall Experience model: Baseline (Perceived Atmosphere + Trust) Nakagawa $R^2_m = 0.637$; adding z-scored Pleasure and Arousal $\rightarrow R^2_m = 0.635$ ($\Delta = -0.002$).\\
Trust model: Baseline (Perceived Atmosphere + Sensory Congruency) $\rightarrow$ adding z-scored Pleasure and Arousal: $\Delta R^2_m = -0.002$; joint Wald for both affect terms $= 0$, $p = 1.00$.\\
Perceived Atmosphere reliably increases Pleasure and (within this dataset) is associated with reduced Arousal; however, once evaluative appraisals enter, affect terms add no marginal explanatory power for Overall Experience or Trust.

\subsection{Assumptions, robustness, and small-sample considerations}\label{supp:S6}

\paragraph{Sphericity \& corrections.} Omnibus rm-ANOVAs applied Greenhouse--Geisser corrections where $\varepsilon<1$; post-hoc contrasts used Bonferroni adjustment (see Supplement~\ref{supp:S1}).

\paragraph{Centering \& random effects.} All endogenous variables were person-mean centered. Under strict within-centering in a balanced 9-cell design, random-intercept fits were singular---expected when between-person variance is removed. Inferences coincide with participant fixed-effects.

\paragraph{Why not confirmatory SEM/CFA in main text?} With $N=24$, confirmatory latent modeling is under-powered and RMSEA is known to be upward-biased in small samples with few indicators. We therefore base inference on rm-ANOVAs and within-person regressions; SEM is reported only as a coherence/visualization check (Supplement~\ref{supp:S2}).

\paragraph{Convergence across pipelines.} The R/\texttt{afex} + \texttt{emmeans} pipeline for rm-ANOVA and post-hoc contrasts underlies the main-text Results. Mixed-model and fixed-effects estimates align numerically; bootstrap mediation results (Supplement~\ref{supp:S4}) align in direction and magnitude with SEM bootstraps (Supplement~\ref{supp:S2}).

\bibliographystyle{unsrt} 
\bibliography{references}

\end{document}